\documentclass[aps,pra,twocolumn,showpacs,superscriptaddress]{revtex4} 
\usepackage{graphicx}
\usepackage{amsmath}
\usepackage{amssymb}
\usepackage{epstopdf}
\usepackage{amsfonts}
\usepackage{dcolumn}
\usepackage{bigints}
\usepackage{xcolor}
\begin{document}
\title{Rabi oscillations and Ramsey-type pulses in ultracold
bosons:
Role of interactions}
\author{Q. Guan}
\address{Homer L. Dodge Department of Physics and Astronomy,
  The University of Oklahoma,
  440 W. Brooks Street,
  Norman,
Oklahoma 73019, USA}
\address{Center for Quantum Research and Technology,
  The University of Oklahoma,
  440 W. Brooks Street,
  Norman,
Oklahoma 73019, USA}
\author{T.~M. Bersano}
\address{Department of Physics and Astronomy, Washington State
University, Pullman, Washington 99164-2814, USA}
\author{S. Mossman}
\address{Department of Physics and Astronomy, Washington State
University, Pullman, Washington 99164-2814, USA}
\author{P. Engels}
\address{Department of Physics and Astronomy, Washington State
University, Pullman, Washington 99164-2814, USA}
\author{D. Blume}
\address{Homer L. Dodge Department of Physics and Astronomy,
  The University of Oklahoma,
  440 W. Brooks Street,
  Norman,
Oklahoma 73019, USA}
\address{Center for Quantum Research and Technology,
  The University of Oklahoma,
  440 W. Brooks Street,
  Norman,
Oklahoma 73019, USA}
\date{\today}

\begin{abstract}
Double-well systems loaded with one, two, or many quantum particles
give rise to intriguing dynamics, ranging from Josephson
oscillation to self-trapping. 
This work presents theoretical and experimental results for
two distinct double-well systems, both created
using dilute rubidium Bose-Einstein condensates with particular
emphasis placed on the role of interaction in the systems.
The first is realized by creating an effective two-level system
through Raman coupling of hyperfine states.
The second is an effective two-level system in momentum space
generated through the coupling by an optical lattice.
Even though the non-interacting systems can,
for a wide parameter range, be described by the same model
Hamiltonian, the dynamics for these two
realizations differ in the presence of interactions.
The difference is attributed to 
scattering diagrams that contribute in the lattice coupled
system but vanish in the Raman coupled system.
The internal dynamics of the Bose-Einstein condensates for both
coupling scenarios is 
probed through a Ramsey-type interference pulse sequence,
which constitutes a key building block of atom interferometers.
These results have important implications in a variety of contexts including lattice calibration
experiments
and momentum space lattices used for quantum analog 
simulations.
\end{abstract}
\maketitle

\section{Introduction}
\label{sec_introduction}

We consider the famous model in which two isolated states are resonantly coupled 
by a monochromatic field, e.g., two energetically separated atomic states
 that are coupled 
by an oscillating electric field.
Starting with all particles populating one of the states,
the population oscillates periodically
between the two states under the influence of the 
external field.
In the presence of a positive or negative detuning $\delta$,
the population still oscillates back and forth;
however, the oscillation period and the maximum transfer probability
(amplitude) of these Rabi oscillations are modified.
This coupled two-level system, which finds
applications in many areas of physics, is discussed in nearly every
quantum text book~\cite{quantumtext,eberly,rabi1939}.
For an interacting ensemble of 
particles,
the Rabi oscillations are, in general, further
modified. In particular,
the population oscillations may not be fully periodic
and the amplitude of the oscillations may decrease or drift
(dephase) due to many-body effects~\cite{saffman2008,gambarelli2012,chen2019}.

This work considers Rabi oscillations in the
context of ultracold atoms,
specifically a degenerate $^{87}$Rb Bose-Einstein condensate
(BEC).
The two-level system is realized 
in two different ways.
In scenario~1,
one-dimensional Raman coupling along the $z$-direction,
realized using two Raman lasers, generates an effective 
pseudo-spin-1/2 system of two
coupled internal
hyperfine states~\cite{spielman2011,spielman2013,zhai2015}.
In scenario~2, a one-dimensional moving optical lattice
along the $z$-direction couples momentum states with
momenta of $2 n \hbar k_L$, where $n$ is an integer and $k_L$ the lattice wave 
vector~\cite{salomon1996,phillips1999,morsch2001}. 
Considering only
the $n=0$ and $n=1$ states, the lattice Rabi coupling case can,
in the absence of interactions, be mapped to the 
same two-state description as the Raman coupling case considered in
scenario~1, 
with
the velocity of the moving lattice 
determining
 the effective detuning.

In the presence of atom-atom interactions,
the Rabi oscillations for scenarios 1 and 2 are found to differ.
The reason for this is traced back to
how the two-level systems are realized.
In second quantization, the interaction potential
takes the form
\begin{eqnarray}
\hat{v} = 
 \int 
\hat{\Psi}^{\dagger}({\bf{r}}_1) \hat{\Psi}^{\dagger}({\bf{r}}_2)
V({\bf{r}}_1,{\bf{r}}_2) 
\hat{\Psi}({\bf{r}}_1) \hat{\Psi}({\bf{r}}_2) d {\bf{r}}_1 d {\bf{r}}_2,
\end{eqnarray}
where 
$\hat{\Psi}^{\dagger}({\bf{r}})$ creates a particle 
at position ${\bf{r}}$. 
Parametrizing $V({\bf{r}}_1,{\bf{r}}_2)$
in terms of a contact interaction of strength $g$,
$V({\bf{r}}_1,{\bf{r}}_2)=
g \delta( {\bf{r}}_1 - {\bf{r}}_2)$,
and
assuming that the field operator $\hat{\Psi}^{\dagger}({\bf{r}})$
can be expressed
in terms of two
states,
\begin{eqnarray}
\hat{\Psi}^{\dagger}({\bf{r}}) = \hat{c}_a^{\dagger} 
\Psi_a^*({\bf{r}})
+
\hat{c}_b^{\dagger} 
\Psi_b^* ( {\bf{r}})
\end{eqnarray}
 ($\hat{c}_a^{\dagger}$ and $\hat{c}_b^{\dagger}$ create
a particle in states $\Psi_a$  and $ \Psi_b$, respectively),
it can be seen that $\hat{v}$ contains 16 terms.
Some of these vanish in the 
Raman coupling case but contribute
appreciably in the  lattice coupling case.

\begin{figure}
\vspace*{-.25in}
\includegraphics[trim=1cm 3cm 0 3cm, width=0.45\textwidth]{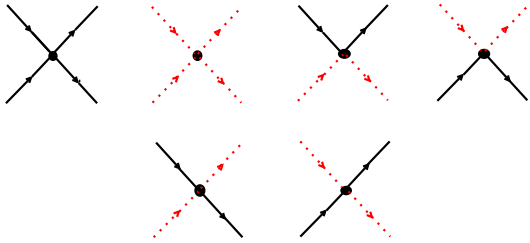}
\vspace*{-.3in}
\caption{Scattering diagrams.
The scattering diagrams in the first row contribute in the
Raman and lattice coupling cases.
The scattering diagrams in the second row vanish in the 
Raman coupling case.
In the first diagram of the first row, particle 1 (top incoming arrow)  and 
particle 2 (bottom incoming arrow) are both in the same
state (both arrows are black solid); the scattering process does not change the state
(the arrows are black solid after the scattering has occured).
In the first diagram of the second row, particle 1 (top incoming arrow)  and 
particle 2 (bottom incoming arrow) are in different
states (black solid and red dotted arrows, respectively); 
after the scattering process, the states of particle 1 (top outgoing arrow)
and
particle 2 (bottom outgoing arrow)  are changed
(red dotted and black solid arrows, respectively).
}
\label{fig_collision}
\end{figure}

We 
write
$\Psi_a({\bf{r}}) = \psi_a ({\bf{r}}) \lvert \uparrow \rangle$
and
$\Psi_b({\bf{r}}) = \psi_b ({\bf{r}}) \lvert \downarrow \rangle$.
In the Raman coupling case, 
$\lvert \uparrow \rangle$
and 
$\lvert \downarrow \rangle$ represent two different hyperfine
states and 
$\psi_a({\bf{r}})$ and $\psi_b({\bf{r}})$
correspond
to
states
with momentum $0$ and $2 \hbar k_R$
($\hbar k_R$ is the Raman momentum).
The resulting non-vanishing interaction terms or scattering diagrams are shown in the first row of
Fig.~\ref{fig_collision}.
The first scattering diagram is proportional to
$\hat{c}^{\dagger}_a \hat{c}^{\dagger}_a \hat{c}_a \hat{c}_a$,
the second scattering diagram
is proportional to $\hat{c}^{\dagger}_b \hat{c}^{\dagger}_b \hat{c}_b \hat{c}_b$,
and the third and fourth scattering diagrams 
are proportional to
$\hat{c}^{\dagger}_a \hat{c}^{\dagger}_b \hat{c}_a \hat{c}_b$.
The last two processes 
can be written as
$\lvert  \Psi_a \rangle_1 + \lvert \Psi_b \rangle_2 \rightarrow \lvert  \Psi_a \rangle_1 + \lvert  \Psi_b \rangle_2$ and
$\lvert  \Psi_b \rangle_1 + \lvert \Psi_a \rangle_2 \rightarrow \lvert  \Psi_b \rangle_1 + \lvert  \Psi_a \rangle_2$,
where the notation $\lvert  \Psi_a \rangle_1$ means that particle~1
occupies state $\Psi_a$.
The  \textquotedblleft mixed scattering diagrams" 
that are
shown in the second row of Fig.~\ref{fig_collision}, which 
are also proportional to $\hat{c}^{\dagger}_a \hat{c}^{\dagger}_b \hat{c}_a \hat{c}_b$,
vanish in the Raman coupling case due to the orthogonality of the
two hyperfine states $\lvert  \uparrow \rangle$ and $\lvert  \downarrow \rangle$.
They correspond to the processes
$\lvert  \Psi_a \rangle_1 + \lvert \Psi_b \rangle_2 \rightarrow \lvert  \Psi_b \rangle_1 + \lvert \Psi_a
 \rangle_2$
and
$\lvert  \Psi_b \rangle_1 + \lvert \Psi_a \rangle_2 \rightarrow \lvert  \Psi_a \rangle_1 + \lvert \Psi_b 
\rangle_2$.

In the lattice system, all the atoms occupy the same hyperfine
state
and
$\Psi_a( {\bf{r}})$ and $ \Psi_b( {\bf{r}})$ 
correspond to 
states
with momentum $\approx 0$ and $\approx 2 \hbar k_L$, respectively.
In this case, $\lvert \uparrow \rangle$
and 
$\lvert \downarrow \rangle$ represent two different plane wave states
and $\Psi_a( {\bf{r}})$ and $ \Psi_b( {\bf{r}})$ 
are not orthogonal to each other. 
As a consequence, the scattering diagrams in the second row of 
Fig.~\ref{fig_collision}
are finite, leading to an enhancemenet of the interaction terms that are
proportional to
$\hat{c}^{\dagger}_a \hat{c}^{\dagger}_b \hat{c}_a \hat{c}_b$.
This factor of two enhancement can be thought of as being due to an exchange process;
it is not specific to bosons and also exists in fermionic systems.
The doubling of the mixed scattering 
diagrams for the lattice coupling case 
compared to the Raman coupling case leads---in 
certain parameter regimes---to distinct Rabi oscillations
for scenarios 1 and 2.
Good agreement between experimental and theoretical results
is found and implications for, e.g., lattice calibration
experiments are discussed.

Working in a parameter regime where the Rabi 
oscillations are noticeably impacted by the interactions,
we probe the internal dynamics of 
condensed atom clouds through a 
Ramsey-type
$\pi/2\text{---hold---}\pi/2$ pulse sequence~\cite{ramsey1985}, 
which is of direct relevance to atom interferometry applications.
Expanding upon earlier work~\cite{castin1996,gupta2011,other_castin}, 
a theoretical framework
is developed that explains the experimentally observed
interference fringes.
$\pi/2$-pulses play an important role in 
momentum space 
engineering~\cite{edwards2010,gadway2015}. 
For example,
the splitting of an initial wave packet
as well as the preparation of a variety of
target states can be accomplished by $\pi/2$ pulses.
Perfect splitting may, however, be hampered by interactions and 
wave packet broadening.
Intriguingly, combining Raman and lattice coupling schemes, 
which are considered separately in this paper, one might
be able to
realize loop-like plaquette systems. 
Specifically, the work presented in this paper provides a stepping stone
for realizing
a plaquette system, in which two momentum space lattices,
occupied by two different hyperfine states, are coupled by Raman lasers.

The remainder of this article is organized as follows.
Section~\ref{sec_experiment} discusses the experimental setup
for the Raman and lattice coupling schemes.
Section~\ref{sec_theory_raman}
presents theory background and numerical results for the Raman
coupling case. It also develops a fully analytical 
framework for the Ramsey-type pulse sequence.
 The agreement between the
theoretical and experimental data for
the Rabi oscillations is excellent.
Section~\ref{sec_theory_lattice} discusses
the lattice coupling case; particular emphasis is placed on
contrasting the dynamics for the lattice coupling scenario with
that for the Raman coupling scenario.
Experimental results are found to agree 
with the
theoretical predictions
quantitatively for the Rabi oscillation data and
qualitatively for the Ramsey-type pulse-sequence data.
Finally, Sec.~\ref{sec_outlook}
summarizes our key findings and presents an outlook.

\section{Experimental setup}
\label{sec_experiment}
The experiments are performed with a $^{87}$Rb BEC consisting of
approximately $N=10^5$  atoms.
Nearly pure BECs are confined in
an elongated harmonic trap with trap frequencies of approximately 
$\{\omega_x, \omega_y,\omega_z\}=2\pi\{140,160,25\}$ Hz. 
The spin-independent trapping potential is produced by two crossed, optical dipole beams with a wavelength of 1064 nm. 
Anharmonic corrections for this trapping configuration are estimated to be negligible for the purpose of this work.
After preparation of the initial state, we remove the trapping potential at time $t=0$.
For all practical purposes, the turning off of the external
confinement is done instantaneously,
\begin{eqnarray}
V_{\text{trap}}({\bf{r}}, t) = 
\left \{
\begin{array}{ll}
\frac{m}{2} \left(\omega_{x}^2x^2+ \omega_y^2 y^2 + \omega_z^2z^2 \right)
& \mbox{ for } t < 0 \\
0 & \mbox{ for } t \ge 0
\end{array}
\right .
,
\end{eqnarray} 
where $m$ denotes the atom mass.
The trap frequencies and atom number are calibrated daily by fitting
dipole oscillation data and cloud 
widths
during expansion, respectively.
The relevant values are reported in the figure captions.

Scenario~1 is realized by applying two 789.1~nm
Raman lasers with effective coupling strength $\Omega_R$
and Raman detuning $\delta_R$
to couple the 
$\lvert F, m_F \rangle=\vert 1,-1 \rangle= \lvert \uparrow \rangle$ 
and
$\lvert F, m_F \rangle=\lvert 1,0 \rangle = \lvert  \downarrow \rangle$ 
hyperfine states of $^{87}\text{Rb}$ under an external magnetic field of approximately 10 G.
Here, $F$ denotes
the total angular momentum of the $^{87}$Rb atom and $m_F$ the
corresponding 
projection quantum number.
The two-photon Raman coupling scheme follows the procedure described in 
Ref.~\cite{experiment_soc}.
In momentum space, the two hyperfine states are separated by
$2 \hbar k_R$, where $k_R$ is determined by the
wave number and orientation of the Raman lasers.
Specifically, the two Raman lasers
with wave vectors
${\bf{k}}_1$ and ${\bf{k}}_2$ cross at an angle of
$\theta_R$.
Defining $2 k_R= | {\bf{k}}_1 - {\bf{k}}_2|$
and using $|{\bf{k}}_1|=|{\bf{k}}_2|$, we
have
$k_R = |{\bf{k}}_1| \sin(\theta_R/2)$;
in our set-up,
$\theta_R \approx \pi/2$
or $k_R \approx |{\bf{k}}_1|/\sqrt{2}$.
The difference between the 
angular frequencies $\omega_1$ and $\omega_2$ of the two lasers 
allows one to set the Raman detuning $\delta_R$,
$\delta_R = 4 E_R - \hbar \omega_R+E_{\text{Zeeman}}$,
where  $ \omega_R= \omega_1 - \omega_2$
and
\begin{eqnarray}
\label{eq_define_er}
E_R = \frac{\hbar^2 k_R^2}{2 m}.
\end{eqnarray}
Here,
$E_{\text{Zeeman}}$ is the Zeeman splitting between
the two coupled hyperfine states. 
The hyperfine state $\lvert  1,1 \rangle$, which is off-resonant 
due to the quadratic Zeeman shift, is not included in our 
theoretical description. We have checked that inclusion of this state
does not notably change the dynamics in the parameter
regime of interest.

Scenario~2 is realized by preparing all atoms in the
$\lvert F,m_F \rangle = \lvert  1,-1 \rangle = \lvert  \uparrow \rangle$
state and loading the BEC into a moving optical lattice~\cite{experiment_optical_lattice}.
Spin changing collisions play a negligible role  in the magnetic fields  used in this work.
The lattice is created by crossing two 1064 nm lasers
at an angle of $\theta_L$ ($\theta_L \approx \pi/2$), 
with polarization perpendicular to the trapping beams, wave vectors ${\bf{k}}_1$ and ${\bf{k}}_2$
($| {\bf{k}}_1|=|{\bf{k}}_2|$), and angular frequencies $\omega_1$ and $\omega_2$.
The resulting lattice is characterized
by the effective
coupling strength $\Omega_L$, the wave vector $k_L$,
and the detuning $\delta_L$
($k_L \approx |{\bf{k}}_1|/\sqrt{2}$
and $\delta_L=4 E_L - \hbar \omega_L$,
where $\omega_L= \omega_1 - \omega_2$).
Energies and lengths are measured in units of
$E_L$ [Eq.~(\ref{eq_define_er}) with the subscript ``$R$" replaced
by ``$L$"] and $1/k_L$, respectively.
Specific values are given in the context of the experiments described below.
In all cases, the external harmonic confinement
is turned off time $t=0$.

In the remainder of this paper,
we denote the coupling strength by $\Omega(t)$ when the 
discussion is independent of the specific scheme, i.e., when the discussion
applies to both the Raman and lattice coupling cases.
When the discussion is specific to one of the scenarios,
we use, respectively, $\Omega_R(t)$ and $\Omega_L(t)$
for the Raman and lattice coupling cases
[correspondingly, $\Omega_0$ in Eqs.~(\ref{eqn8}) and (\ref{eqn7})
below
are replaced by $\Omega_{0,R}$ and $\Omega_{0,L}$, respectively].
The coupling, which is assumed to be real, is turned on at time 
$t_{\text{start}}$, where $t=0$ is the time at which the trapping potential is removed.
For $t_{\text{start}}>0$, the initial 
BEC
expands in the absence of the Raman or lattice
drive,
thereby reducing the interaction
strength during the subsequent pulse sequence.
For the Rabi oscillation measurements, we keep
$\Omega(t)$ on for a time interval 
$t_{\text{seq}}=t_{\text{end}}-t_{\text{start}}$,
\begin{eqnarray}
\label{eqn8}
\Omega(t) = 
\left \{
\begin{array}{ll}
0  & \text{ for } t < t_{\text{start}}  \\
\Omega_0 & \text{ for } t_{\text{start}} \le t < t_{\text{end}}\\
0 &  \text{ for } t \ge t_{\text{end}} 
\end{array}
\right .
.
\end{eqnarray} 
For the Ramsey-type pulse sequence of length $t_{\text{seq}}=\tau_1+t_{\text{hold}}+\tau_2$, 
the coupling strength $\Omega(t)$ reads
\begin{eqnarray}
\label{eqn7}
\Omega(t)=
\left \{
\begin{array}{ll}
0 & \mbox{ for } t < t_{\text{start}} \\
\Omega_0 & \mbox{ for } t_{\text{start}} \le t < t_{\text{start}} + \tau_1 \\
0 & \mbox{ for } t_{\text{start}} + \tau_1 \le t < t_{\text{start}} + \tau_1 + t_{\text{hold}} \\
\Omega_0 & \mbox{ for }  t_{\text{start}} + \tau_1 + t_{\text{hold}} \le t < t_{\text{end}} \\
0 & \mbox{ for } t \ge t_{\text{end}}\\
\end{array}
\right .
.
\end{eqnarray}
In the experiment, the turning on of the coupling strength
is not quite instantaneous but instead occurs over about $75$~$\mu$s.
To facilitate the comparison 
between theory and experiment,
we choose $t_{\text{start}}$ to be the time at which
$\Omega(t)$ has reached half of its maximum. 
In many applications that involve momentum transfer, 
a  $\pi/2$-wait-$\pi$-wait-$\pi/2$
pulse sequence is used instead of the shorter
$\pi/2$-wait-$\pi/2$ pulse sequence. The reason we decided to
apply the simpler pulse sequence is that the ``$\pi$ reversal pulse" does not, as in other
scenarios, remove the linear phase in our systems due to the
presence of interactions.

The imaging is done at time $t_{\text{end}}+t_{\text{ToF}}$, i.e., after 
an additional expansion time of $t_{\text{ToF}}$.
In the absence of the trapping potential, the momentum components separate 
naturally due to the fact that
the states $\Psi_a$ and $\Psi_b$ have different velocities.
For the lattice case, e.g.,
an expansion time of $t_{\text{ToF}} \approx 10$~ms
corresponds to a
separation of the cloud centers by
about 
$85$~$\mu$m
along the $z$-direction.
This distance is larger than the size of the clouds after the
expansion. For an initial
cloud with Thomas-Fermi radius $22$~$\mu$m, e.g., the size
of the cloud at time $t_{\text{seq}}+t_{\text{ToF}}$
is about $43$~$\mu$m.

Depending on the observable, the time-of-flight expansion plays no role, a negligible role,
or an essential role when comparing experimental and
theoretical data.
For the Raman coupling case, the populations of the states
$\Psi_a$ and $\Psi_b$ do not change during the time-of-flight
expansion. 
This implies that
theoretical results for the populations, calculated by neglecting the 
time-of-flight expansion, can 
be compared directly with experimentally
measured populations. 
Correspondingly, we do not simulate the time-of-flight 
sequence when we compare Rabi oscillation data.
Experiment-theory comparisons of the Ramsey-type 
pulse sequence,
in contrast, require that the
time-of-flight expansion be simulated to explain the observed fringe structures.

For the lattice case, the situation
is slightly different.
The populations of the states $\Psi_a$ and $\Psi_b$, which are distinguished only by their momentum, can change
during the time-of-flight expansion due to atom-atom
collisions
that
involve states
with momenta $\approx n \hbar k_L$, where
$n = -2, \pm 4,\pm 6,\cdots$.
However, such population transfer is typically small;
note that this is the reason why
the 
two-state model introduced in
Sec.~\ref{two_momentum_component} 
provides a reliable description for a fairly large parameter window.
The small population transfer
implies that the dynamics during the time-of-flight expansion
can, in a first approximation, be neglected when analyzing populations.
Understanding the 
internal dynamics such as the formation of density patterns, in contrast, requires that the time-of-flight sequence be modeled explicitly.

\section{Raman coupling case}
\label{sec_theory_raman}

\subsection{General framework}
\label{sec_general_framework}
Our theoretical analysis of the Raman-coupled system is
based on the standard
mean-field formulation~\cite{zhai2015}, 
which writes the mean-field spinor in terms of the
components $\psi_a({\bf{r}},t)$ and $\psi_b({\bf{r}},t)$.
Here and in what follows, $\psi_a$ and $\psi_b$ are 
time-dependent; 
note that the discussion in Sec.~\ref{sec_introduction}
adopted a stationary framework for simplicity.
The unrotated $2\times 2$ mean-field Hamiltonian $\hat{H}$ reads 
\begin{widetext}
\begin{eqnarray}
\label{eqn1}
\hat{H}= 
&&
\left(\frac{\hat{\bf{p}}^2}{2m}+V_{\text{trap}}({\bf{r}}, t)\right)\otimes I_2 
 + 
\begin{pmatrix}
  &g_{aa}|\psi_{a}({\bf{r}},t)|^2+g_{ab}|\psi_{b}({\bf{r}},t)|^2 & 0\\
  & 0 & g_{ba}|\psi_{a}({\bf{r}},t)|^2+g_{bb}|\psi_{b}({\bf{r}},t)|^2\\
\end{pmatrix} 
 + \nonumber \\
&& 
\begin{pmatrix}
  &0 & \frac{\Omega_R(t)}{2}\exp(-2\imath k_R z+ \imath \omega_R t)\\
  &\frac{\Omega_R(t)}{2}\exp(2\imath k_R z- \imath \omega_R t) & E_{\text{Zeeman}}
\end{pmatrix},
\end{eqnarray}
\end{widetext}
where $I_{2}$ is the $2\times 2$ identity matrix
and the normalization, expressed in terms
of the fractional populations $N_a$ and $N_b$, is
\begin{eqnarray}
\label{eq_norm1}
N_a+N_b=1
\end{eqnarray}
with ($j=a$ or $b$)
\begin{eqnarray}
\label{eq_norm2}
N_j=\int  |\psi_j({\bf{r}},t)|^2   d {\bf{r}}.
\end{eqnarray}
The 
interaction strengths
$g_{ij}$  between atoms in hyperfine states $i$ and $j$
are given by
\begin{eqnarray}
g_{ij} = \frac{4 \pi \hbar^2 (N-1)a_{ij}}{m}.
\end{eqnarray}
For $^{87}$Rb, we
have
$a_{aa} = 100.4$~$a_{\text{B}}$,
$a_{ab} = a_{ba}=100.4$~$a_{\text{B}}$, and
$a_{bb} = 100.9$~$a_{\text{B}}$~\cite{scatteringlength_rb}, 
where $a_{\text{B}}$ denotes the Bohr radius.
In the arguments presented in Sec.~\ref{sec_introduction}, the four $g_{ij}$ were 
assumed to be the same; this simplifying assumption
is again made in Sec.~\ref{sec_theory_ramsey_ana}.
The time dynamics of the system is governed by 
\begin{eqnarray}
\label{eqn2}
\imath\hbar\frac{\partial}{\partial t} 
\begin{pmatrix}
\psi_a({\bf{r}}, t) \\
\psi_b({\bf{r}}, t)
\end{pmatrix}
= \hat{H} 
\begin{pmatrix}
\psi_a({\bf{r}}, t) \\
\psi_b({\bf{r}}, t)
\end{pmatrix}.
\end{eqnarray}
Defining the rotated states 
$\tilde{\psi}_a({\bf{r}},t)$ and $\tilde{\psi}_b({\bf{r}},t)$,
\begin{eqnarray}
\label{eqn3}
\begin{pmatrix}
\tilde{\psi}_a({\bf{r}}, t) \\
\tilde{\psi}_b({\bf{r}}, t)
\end{pmatrix}
=
\hat{U}(z, t)
\begin{pmatrix}
\psi_a({\bf{r}}, t) \\
\psi_b({\bf{r}}, t)
\end{pmatrix},
\end{eqnarray}
in terms of the rotation operator $\hat{U}(z, t)$,
\begin{eqnarray}
\label{eqn4}
\hat{U}(z, t)=
\begin{pmatrix}
  &1 & 0\\
  &0 &\exp(-2\imath k_R z + \imath \omega_R t)
\end{pmatrix},
\end{eqnarray}
we obtain
Eq.~(\ref{eqn2}) with
${\psi}_a({\bf{r}}, t)$, ${\psi}_b({\bf{r}}, t)$, and $\hat{H}$
replaced by
$\tilde{\psi}_a({\bf{r}}, t)$, $\tilde{\psi}_b({\bf{r}}, t)$, and $\hat{\tilde{H}}$,
respectively,
where the rotated Hamiltonian $\hat{\tilde{H}}$ is given by
\begin{widetext}
\begin{eqnarray}
\label{eqn6}
\hat{\tilde{H}}= &&
\left(\frac{\hat{\bf{p}}^2}{2m}+V_{\text{trap}}({\bf{r}}, t)\right)\otimes I_2 
 +
\begin{pmatrix}
  &g_{aa}|\tilde{\psi}_{a}({\bf{r}},t)|^2+g_{ab}|\tilde{\psi}_{b}({\bf{r}},t)|^2 & 0\\
  & 0 & g_{ba}|\tilde{\psi}_{a}({\bf{r}},t)|^2+g_{bb}|\tilde{\psi}_{b}({\bf{r}},t)|^2\\
\end{pmatrix} + \\\nonumber
&& 
\begin{pmatrix}
  &0 & \frac{\Omega_R(t)}{2}\\
  &\frac{\Omega_R(t)}{2}& \frac{2\hbar k_R \hat{p}_z}{m} + \delta_R
\end{pmatrix}. 
\end{eqnarray}
\end{widetext}
To obtain Eq.~\eqref{eqn6},
we used the relation 
$|\tilde{\psi}_{j}({\bf{r}}, t)|^2=|\psi_{j}({\bf{r}}, t)|^2$,
where $j=a$ or $b$.
Importantly, the position- and time-dependent
phase $\tilde{\gamma}_b({\bf{r}},t)$
of the rotated component
$\tilde{\psi}_{b}({\bf{r}}, t)$
differs from the phase ${\gamma}_b({\bf{r}},t)$
of the unrotated component ${\psi}_{b}({\bf{r}}, t)$.
Since the change of the phases of the unrotated spinor components is dominated by
the laser coupling term, thereby masking the change due to the internal dynamics,
it is more convenient to analyze the phases of the spinor-components 
in the rotated basis, whose phase dynamics is 
governed by
\textquotedblleft internal effects''
as opposed to the laser coupling.

For the Rabi oscillation measurements and the Ramsey-type pulse sequence, 
the BEC is initially (i.e., at $t=0$) prepared in the  state 
$\psi_a({\bf{r}},t)\lvert \uparrow \rangle = \tilde{\psi}_a({\bf{r}},t)\lvert \uparrow \rangle$, which
is characterized by a vanishing
average mechanical momentum along
the $z$-direction, i.e., $\langle \hat{p}_z\rangle_{\text{initial}}=0$. 
Our calculations assume an axially symmetric harmonic trap
with $\omega_x=\omega_y=\omega_{\rho}$.
The trapping potential defines the harmonic oscillator lengths $a_{\text{ho},z}$
and $a_{\text{ho},\rho}$,
\begin{eqnarray}
a_{\text{ho},z/\rho}=\sqrt{\frac{\hbar}{m \omega_{z/\rho}}}.
\end{eqnarray}
The coupled mean-field equations are solved using standard techniques.
The initial state is obtained by imaginary time 
propagation. The real time dynamics is
implemented by expanding the time evolution
operator in terms of Chebychev polynomials~\cite{chebychev1,chebychev2}.
We use equally spaced grid points in $z$ and $\rho$.
The convergence of the results presented
has been tested
with respect to the size of the simulation box, 
the number of grid points,
and the time step.

\subsection{Rabi oscillations: Vanishing Raman detuning}
\label{sec_ramanrabi}

\begin{figure}
\includegraphics[width=0.4\textwidth]{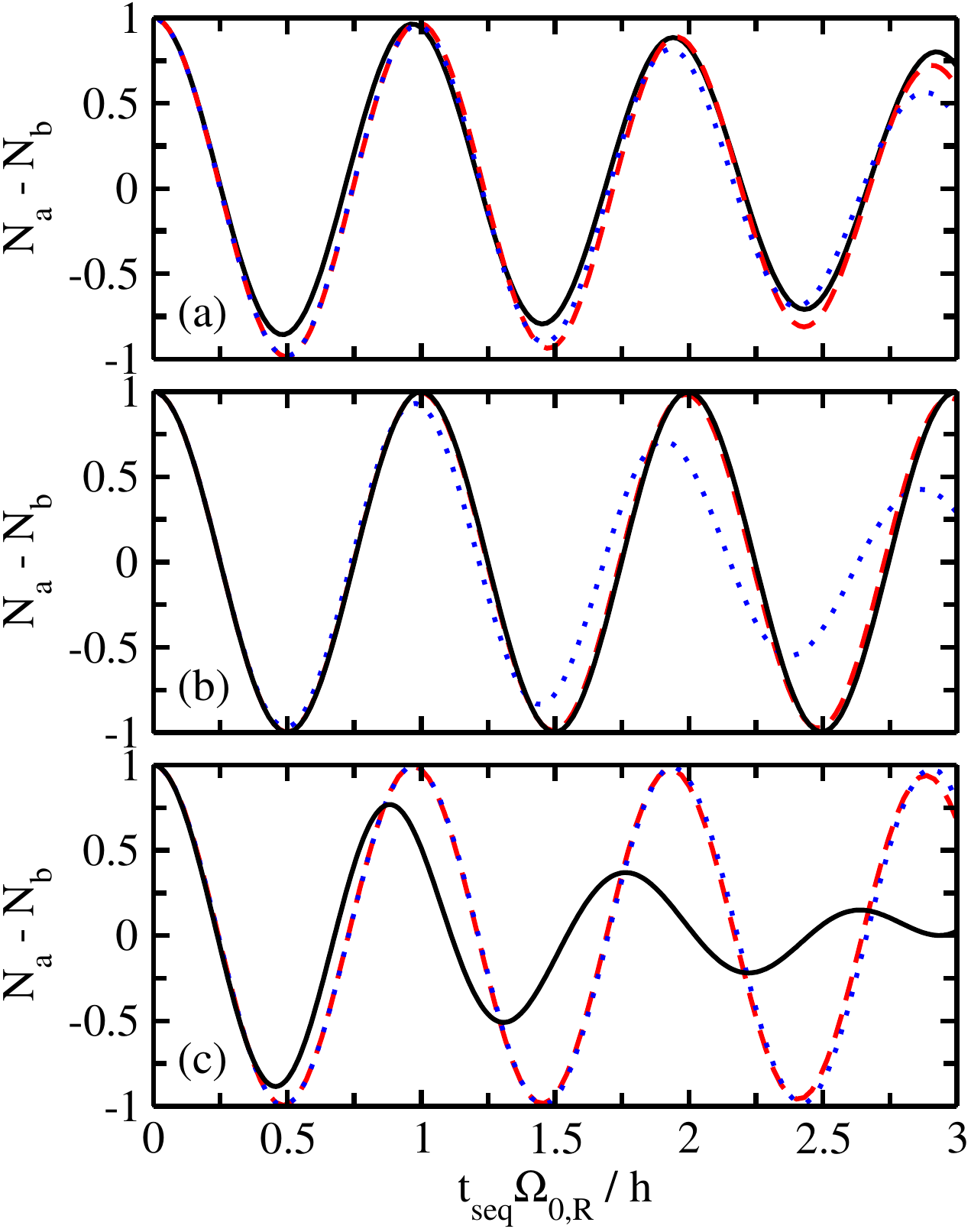}
\caption{Rabi oscillations for Raman coupling case
 (numerical results).
The lines show the difference $N_a-N_b$
between the fractional populations
as a function of the dimensionless time 
$t_{\text{seq}} \Omega_{0,{{R}}}/h$
for
$t_{\text{start}}=0$, $E_R/h=1960$~Hz,
$\delta_R=0$, and $\omega_{\rho}=2 \pi \times 200$~Hz.
(a) Changing the particle number $N$.
The black solid, red dashed, and blue dotted lines 
are for $N=1$, $N=3 \times 10^5$, and $N=10^6$, respectively.
The weak trapping frequency is $\omega_z=2 \pi \times 40$~Hz
and the coupling strength is $\Omega_{0,R}=E_{{R}}$. 
(b) Changing the angular trapping frequency $\omega_z$.
The black solid, red dashed, and blue dotted lines 
are for 
$\omega_z= 2 \pi \times 10$~Hz,
$\omega_z= 2 \pi \times 20$~Hz, and
$\omega_z= 2 \pi \times 60$~Hz, respectively.
The coupling strength is $\Omega_{0,R}=E_{{R}}$ and the number
of particles is $N=3 \times 10^5$. 
(c) Changing the coupling strength $\Omega_{0,{R}}$.
The black solid, red dashed, and blue dotted lines 
are for $\Omega_{0,{R}}=E_{{R}}/2$,
$\Omega_{0,{R}}=3 E_{{R}}/2$,
and
$\Omega_{0,{R}}=5 E_{{R}}/2$, respectively.
The number of particles is $N=3 \times 10^5$ and the weak trapping
frequency is $\omega_z= 2 \pi \times 40$~Hz.
}
\label{fig_ramanrabi}
\end{figure}

This section discusses Rabi oscillation results for the 
Raman coupling case
with $\delta_R=0$. The numerical solutions are obtained
by solving 
the time-dependent mean-field equation
for $\hat{\tilde{H}}$ [see Eq.~(\ref{eqn6})]
with
$E_R/h=1960$~Hz.
The Raman coupling is turned on
at $t=0$, i.e., we have $t_{\text{start}}=0$.
Figs.~\ref{fig_ramanrabi}(a)-\ref{fig_ramanrabi}(c) show the
difference $N_a-N_b$
between the fractional populations
as a function
of the dimensionless time
$t_{\text{seq}}\Omega_{0,R}/h$ for different $N$, $\omega_z$, and $\Omega_{0,R}$,
respectively.

Figure~\ref{fig_ramanrabi}(a) shows numerical results for
$\Omega_{0,R}=E_R$ and 
three different $N$, namely $N=1$,
$3 \times 10^5$, and $N=10^6$.
Even though the Rabi coupling lasers are turned on, at time $t=0$, after the
trapping potential has been switched off, the figure caption quotes the
trapping frequencies since they determine the initial state and thus the distribution
of the kinetic and potential energy, including the mean-field energy, in the system.
For the non-interacting single-atom system
[black solid line in Fig.~\ref{fig_ramanrabi}(a)],
the Rabi oscillation period is nearly constant for the times considered; 
the amplitude, however, is visibly damped.
While this \textquotedblleft non-perfect'' sinusoidal 
behavior might be surprising at first sight,
it can be explained as follows:
The 
center of mass of the component $\tilde{\psi}_b({\bf{r}},t)$
moves relative to the center of mass of the
component $ \tilde{\psi}_a({\bf{r}},t)$ during the
Rabi oscillations. Thus, 
the two components are not
perfectly overlapping spatially.
As a consequence, the relative phase of the spinor components at fixed
${\bf{r}}$ is changing slightly due to 
the relative motion of the two components with respect
to each other. This phase difference is responsible for the
non-perfect population transfer (``damping").
An alternative but equivalent picture is that the finite momentum width of the initial state 
corresponds to a small effective momentum-dependent
detuning.
This effective detuning decreases with increasing mean-field interactions
due to the decrease of the width of the initial state in momentum space.

When mean-field interactions are present
[the red dashed and blue dotted lines in Fig.~\ref{fig_ramanrabi}(a)
are for $N=3 \times 10^5$ and $N=10^6$, respectively], the amplitude 
of the Rabi oscillation data 
changes somewhat while
the oscillation period is essentially unaffected by the
interactions.
For these two $N$-values, the chemical
potential (in units of $h$) of the initial state
is 
$\approx 3303$~Hz and $5347$~Hz
(corresponding to $1.685E_R$ and $2.728E_R$), i.e., the
chemical potential at $t=0$ is larger than $\Omega_{0,R}$.

To highlight the effect of the
interactions, Fig.~\ref{fig_ramanrabi}(b)
shows the Raman-induced Rabi oscillations for $N=3 \times
10^5$ 
[same atom number as used for the red dashed line
in Fig.~\ref{fig_ramanrabi}(a)]
for weaker and stronger confinement along the 
$z$-direction than used in  Fig.~\ref{fig_ramanrabi}(a).
Stronger  
confinement leads to higher density and thus to enhanced interaction effects.
For the largest $\omega_z$ considered [blue dotted
line in Fig.~\ref{fig_ramanrabi}(b)], 
the fractional population difference
$N_a - N_b$
deviates
appreciably
from a simple sinusoidal 
curve
after a few oscillations.
This indicates that care needs to be taken when calibrating the
effective Raman coupling strength
$\Omega_{0,R}$; in particular, a fit to a simple sinusoidal
function, applicable to the ideal two-level model, might not yield the 
correct effective
coupling strength.

The results presented in 
Figs.~\ref{fig_ramanrabi}(a)-\ref{fig_ramanrabi}(b) 
are for $\Omega_{0,R}=E_R$.
Figure~\ref{fig_ramanrabi}(c) 
shows the dynamics for 
smaller and larger coupling strengths, namely $\Omega_{0,R}=E_R/2$,
$\Omega_{0,R}=3E_R/2$, $\Omega_{0,R}=5E_R/2$, and the same trap confinement as in Fig.~\ref{fig_ramanrabi}(a).
Even though the particle number is quite
moderate (namely, $N=3 \times 10^5$),
the oscillation amplitude and period 
for $\Omega_{0,R}=E_R/2$
[solid line in Fig.~\ref{fig_ramanrabi}(c)]
deviate 
strongly from perfect sinusoidal
behavior due to the enhanced effects of the mean-field
energy with decreasing $\Omega_{0,R}$.
This implies that care needs to be exercised if 
the calibration of the effective Raman coupling strength
is done for low coupling strengths. 
In this regime, one has to make sure that 
the particle number is sufficiently low or that one allows for sufficient
time-of-flight expansion prior to 
turning on the Raman Rabi coupling
(Fig.~\ref{fig_ramanrabi} is for $t_{\text{start}}=0$).
If this is not done, the value of the effective
Rabi coupling strength, which enters into the underlying system Hamiltonian,
may be impacted by interaction effects, potentially leading to errors in
experiments that require reliable precision, such as quantum analog simulations
and many-body studies.
Alternatively, 
explicit comparisons with Gross-Pitaevskii equation results, as done in this work, would be
very useful when interactions are present.
Last, one may consider
performing
the calibration 
in the 
\textquotedblleft large power'' regime and 
extrapolating the resulting calibration
curve instead of performing the calibration in the
\textquotedblleft low power'' regime.

\subsection{Rabi oscillations: Theory-experiment comparison}
\label{sec_raman_rabi_experiment}

\begin{figure}
\includegraphics[width=0.4\textwidth]{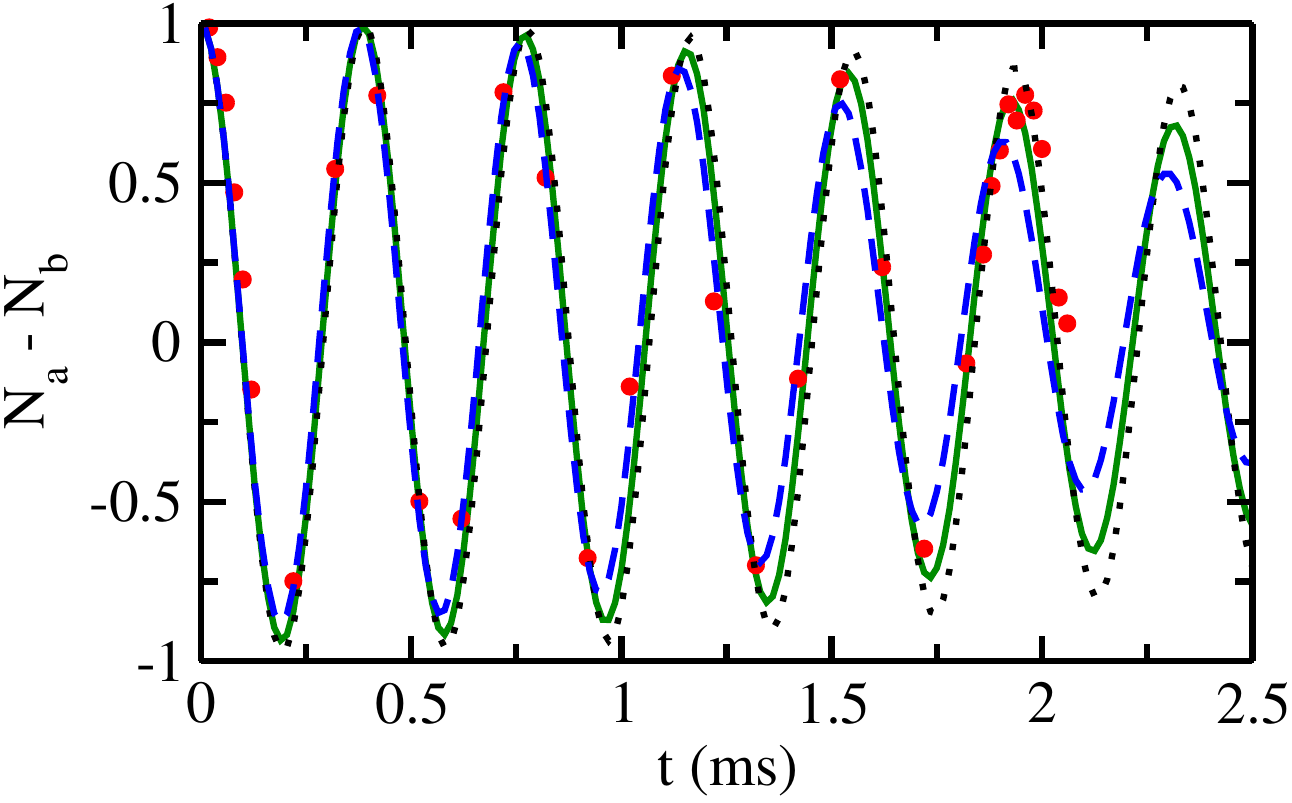}
\caption{Theory-experiment comparison for Raman Rabi oscillations
for a $^{87}$Rb BEC.
The red dots show the experimentally determined 
difference in the fractional populations as a
function of time. The experimental parameters are
$\omega_x = 2 \pi \times 155$~Hz,
$\omega_y = 2 \pi \times 179$~Hz,
$\omega_z = 2 \pi \times 24.8$~Hz,
$N=1.5 \times 10^5$,
$E_R/h=1960$~Hz,
$\Omega_{0,R}/h=2548$~Hz,
and 
$t_{\text{start}}=0.5$~ms.
The experiments were performed for
a small negative detuning
$\delta_R/h$ that is estimated to be between $-200$~Hz and $-600$~Hz, where the 
uncertainty 
is 
due to
fluctuations in the external magnetic field responsible for the Zeeman splitting.
The solutions to the mean-field Gross-Pitaevskii equation (lines)
are obtained for an axially symmetric trap
characterized by the experimentally measured $\omega_z$ and 
$\omega_{\rho}=  2 \pi \times167.0$~Hz
($\omega_{\rho}$ is taken to be the mean of the experimental
$\omega_x$ and $\omega_y$).
The black dotted, green solid, and blue dashed lines show results for
$\delta_R/h=-200$~Hz, $-400$~Hz, and $-600$~Hz, respectively; 
the other parameters are taken from experiment.
The
results for $\delta_R/h=-200$~Hz describe the experimental data the best.
   The chemical potential $\mu$ prior to turning off the trap is
 $0.8436$~$E_R$.
 The mean-field energy per particle  prior to turning off the trap and after the 
 $0.5$~ms expansion is
 $0.5206$~$E_R$ and $0.4076$~$E_R$, respectively.
%The experimental data is "180727" data.
}
\label{fig_ramanrabiexp}
\end{figure}

Figure~\ref{fig_ramanrabiexp} 
shows a comparison between theory and experiment for the Raman-induced Rabi oscillations
for an initial state with chemical potential $\mu=0.8436$~$E_R$.
This chemical potential corresponds to a mean-field energy per particle 
at $t=0$ of $0.5206$~$E_R$. To reduce the mean-field energy in the system,
a $0.5$~ms free expansion step was inserted after tuning off the trap
and prior to turning on the Raman coupling lasers. At the end of the 
free expansion, the mean-field energy per particle is about $20$~\% smaller than at $t=0$.
The resulting Rabi oscillations are slightly damped.
Although the experimental data (red dots) are
obtained for a small negative detuning
$\delta_R$, the detuning is not the only cause for the damping.
Extrapolating the
mean-field Gross-Pitaevskii
results for finite detuning to zero detuning,
we conclude that even the zero detuning case
displays damping [explicit calculations for $\delta_R=0$ (not shown)
confirm this].
Combining the good agreement between the
experimental data and theoretical curves with the discussion
of the previous section, we conclude that the damping 
can be partially attributed to the mean-field interactions. Indeed,
if we let the BEC expand longer prior to turning on the Raman coupling, the 
damping or dephasing, for the same detuning $\delta_R$, is reduced.

Interestingly, fitting of the mean-field
Gross-Pitaevskii results for $\delta_R/h=-200$~Hz 
(this detuning gives the best agreement with the experimental data)
to a damped periodic function of the form
\begin{eqnarray}
\label{eq_fit}
N_a-N_b = \cos( 2\pi f t) \exp(-t/\tau)
\end{eqnarray}
yields a frequency $f$ of $2578$~Hz.
This ``fitted Rabi coupling strength" is about 1.2~\% larger
than the Rabi coupling strength $\Omega_{0,R}$
used in the simulations. This indicates that the 
interactions impact, for the parameter combinations considered, 
the oscillation frequency
much less than the amplitude.
More specifically, for the parameter combination considered
in Fig.~\ref{fig_ramanrabiexp}, the
effect of the interactions on the Rabi oscillations can be 
described, to a good approximation,
phenomenologically by 
the time constant $\tau$.

\subsection{Ramsey-type pulse sequence: Theory overview}
\label{sec_ramanramsey_overview}

Throughout this section, 
the initial BEC ($N=3 \times 10^5$) is prepared
for a confinement with $\omega_{\rho}= 2 \pi \times 200$~Hz
and $\omega_z = 2 \pi \times 40$~Hz.
The dynamics are analyzed for the Ramsey-type pulse sequence
with Raman coupling strength $\Omega_{0,R}=E_R$, 
detuning $\delta_R=0$, and---as in 
Sec.~\ref{sec_ramanrabi}---$E_R/h=1960$~Hz and $t_{\text{start}}= 0$.
The main emphasis lies on developing,
motivated by numerical simulations
of the time-dependent mean-field equation
for the Hamiltonian given in Eq.~(\ref{eqn6}),
a benchmark
and physical picture
that provides the motivation
for the analytical treatment presented in Sec.~\ref{sec_theory_ramsey_ana}.

\begin{figure*}
\includegraphics[width=1.0\textwidth]{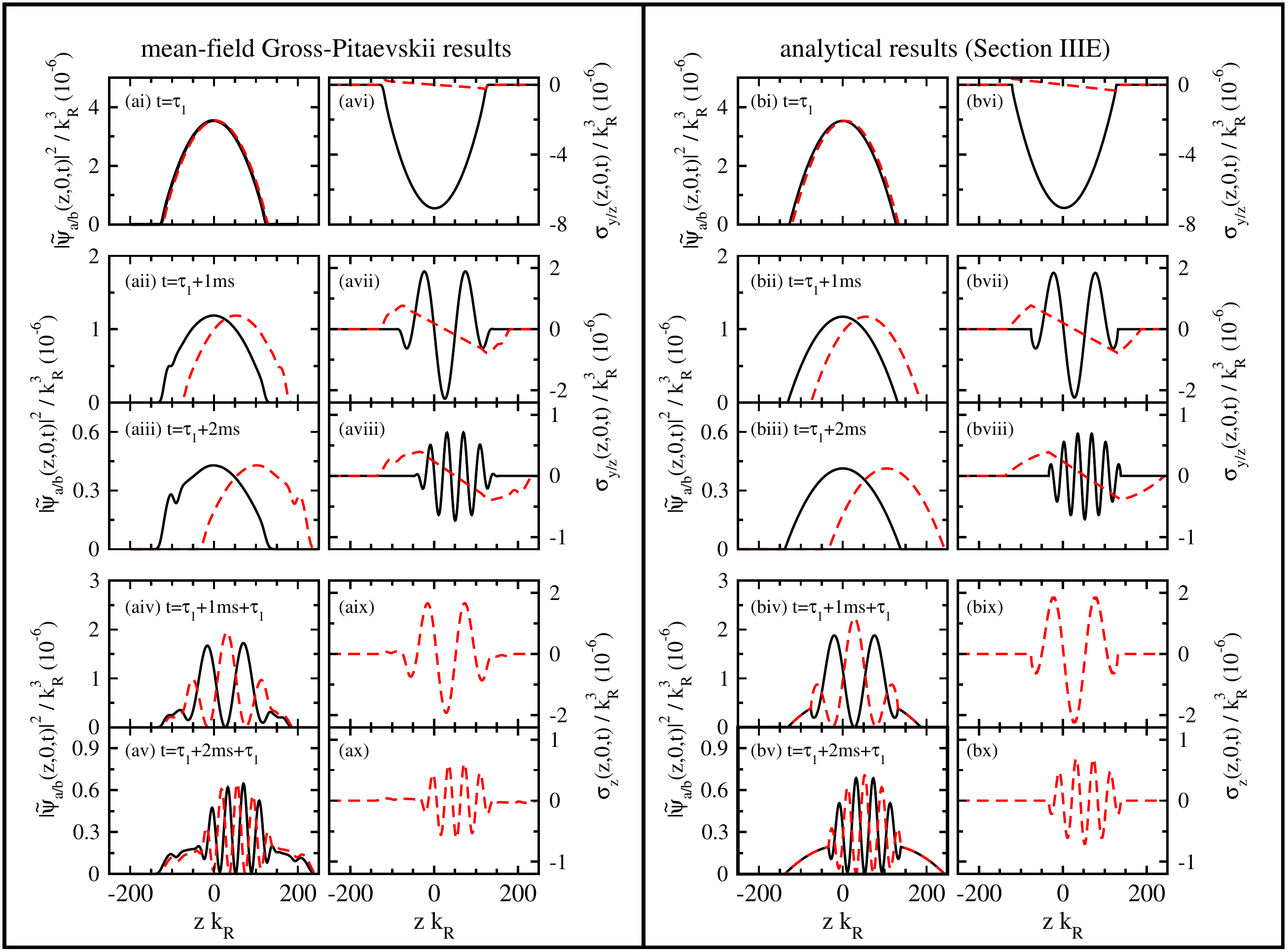}
\caption{Density cuts and local
spin expectation values
for the Ramsey-type pulse sequence 
with $E_R/h=1960$~Hz,  $\Omega_{0,R}=E_R$,
and  $\delta_R=0$ (theory results).
The $^{87}$Rb BEC consists of $N=3 \times 10^5$ atoms 
and is prepared in an
axially symmetric trap with $\omega_{\rho} =2 \pi \times 200$~Hz
and $\omega_z= 2 \pi \times 40$~Hz.
This
corresponds to a chemical potential, in units of $h$,
of $\approx 3303$~Hz. 
All results are obtained for $t_{\text{start}}=0$.
The first and second columns are obtained by solving 
the time-dependent mean-field equation for the
Hamiltonian given in Eq.~(\ref{eqn6}) numerically.
The third and fourth columns show the same observables as 
the first and second columns but are, instead, calculated using the 
fully analytical 
framework developed in Sec.~\protect\ref{sec_theory_ramsey_ana};
the agreement is very good.
The black solid and red dashed lines 
in panels~(ai)-(av)
show the
density cuts $|\tilde{\psi}_a(z,0,t)|^2$ and
$|\tilde{\psi}_b(z,0,t)|^2$, respectively. 
The black solid and red dashed lines 
in panels~(avi)-(ax)
show the
local spin expectation values
$\sigma_y(z,0,t)$ and
$\sigma_z(z,0,t)$,
respectively. 
The time increases from the first row,
to the second/third row, to the fourth/fifth row (the value of the time is given in the panels);
the second and fourth row correspond to a hold time of $1$~ms, and
the third and fifth row correspond to a hold time of $2$~ms. It can be seen that the agreement 
between the mean-field Gross-Pitaevskii results and the fully analytical results
is quite good. The figure illustrates that the second $\pi/2$-pulse transfers the
information encoded in $\sigma_y(z,0,t)$ to
$\sigma_z(z,0,t)$, making the interference visible in the population difference.
}
\label{fig_ramanramsey}
\end{figure*}

When the Raman coupling is turned on,
population is transferred from the component $\tilde{\psi}_a({\bf{r}},t)$
to the component $\tilde{\psi}_b({\bf{r}},t)$.
As discussed in the previous sections, the interactions can notably impact the
Rabi oscillations, in particular for longer times. Despite of this,
we measure the lengths of our pulses in terms of the characteristic
time scale of the non-interacting system, i.e., we refer to a $\pi/2$-pulse as 
a pulse that transfers, in the absence of interactions
and assuming an infinitely narrow momentum space wave
packet, 
half of the atoms from the state $\tilde{\psi}_a({\bf{r}},t)$ to 
the state
$\tilde{\psi}_b({\bf{r}},t)$.
For the parameters employed
in Fig.~\ref{fig_ramanramsey}, a $\pi/2$-pulse corresponds
to $\pi h/(2 \Omega_{0,R})\approx 0.1276$~ms.
Figure~\ref{fig_ramanramsey}(ai) 
shows that the population of state $\tilde{\psi}_b({\bf{r}},t)$
after the first $\pi/2$-pulse is close to $50$~\% (it is $49.95$~\%).
In addition, it can be seen that the population in 
the component 
$\tilde{\psi}_b({\bf{r}},t)$ moves a tiny bit relative to the population in 
the component $\tilde{\psi}_a({\bf{r}},t)$ during the first $\pi/2$-pulse.
The reason is that the population in state $ \tilde{\psi}_b({\bf{r}},t)$ 
has an average mechanical momentum of about $2 \hbar k_R$
along the $z$-direction
while the population in state
$\tilde{\psi}_a({\bf{r}},t)$ has an average mechanical momentum 
very close to zero along the $z$-direction. 

During the variable hold time, 
no population transfer occurs since the Raman coupling lasers
are turned off.
The two key characteristics during the hold time are:
First, the population in state $\tilde{\psi}_b({\bf{r}},t)$ continues to move
relative to that in 
component $\tilde{\psi}_a({\bf{r}},t) $.
Second, the interacting BEC expands a tiny bit. 
Figures~\ref{fig_ramanramsey}(aii) 
and \ref{fig_ramanramsey}(aiii)
show $\rho=0$ cuts and
are for  $t_{\text{hold}}=1$~ms and $t_{\text{hold}}=2$~ms, respectively.

To \textquotedblleft reunite'' the populations of the states $\tilde{\psi}_a({\bf{r}},t)$ and 
$\tilde{\psi}_b({\bf{r}},t)$,
a second $\pi/2$-pulse is applied.
For the example shown in Fig.~\ref{fig_ramanramsey}, the population of 
state $\tilde{\psi}_a({\bf{r}},t)$ after the second 
$\pi/2$-pulse is $51.74$~\% for $t_{\text{hold}}=1$~ms
and $50.15$~\% for $t_{\text{hold}}=2$~ms.
After the second $\pi/2$-pulse, 
the density profiles
[see Figs.~\ref{fig_ramanramsey}(aiv)-\ref{fig_ramanramsey}(av)
for $\rho=0$ cuts]
show interference fringes in the \textquotedblleft central'' or \textquotedblleft overlap''
region, i.e., in the spatial region where the two clouds overlapped
prior to the application of the second $\pi/2$-pulse.
The interference fringes establish themselves through
out-of-phase oscillations of 
$|\tilde{\psi}_a(z,0)|^2$ and $|\tilde{\psi}_b(z,0)|^2$.
We observe analogous fringes for other $\rho$ values.
The total density of the two components (not shown), in contrast,
exhibits  oscillations with comparatively small amplitude in the outer region and
no oscillations in the central region.

Quite generally, the appearance of fringes
such
as those displayed in Figs.~\ref{fig_ramanramsey}(aiv) 
and \ref{fig_ramanramsey}(av) 
suggests
the existence of two interfering pathways,
i.e., the existence of a spatially dependent 
phase difference.
In the following, we introduce a theoretical framework that
highlights
how the fringe pattern develops for the Ramsey-type pulse sequence
with Raman coupling.
To this end,
it is instructive to visualize the time-evolving rotated
spinor on the
Bloch sphere.
Since the two components 
$\tilde{\psi}_a({\bf{r}},t)$ and $\tilde{\psi}_b({\bf{r}},t)$
can each be written in terms of one complex number for each
$z$, $\rho$, and $t$
(the axial symmetry suggests the use of cylindrical coordinates),
we define
\begin{eqnarray}
\label{eqn9}
\begin{pmatrix}
\tilde{\psi}_a(z, \rho, t) \\
\tilde{\psi}_b(z, \rho, t)
\end{pmatrix}
=R(z, \rho, t)\exp \left[\imath\tilde{\gamma}_a(z, \rho, t)\right] \times \nonumber \\
\begin{pmatrix}
\cos\big(\frac{\theta(z, \rho, t)}{2}\big) \\
\exp\left[ \imath\phi(z, \rho, t) \right] \sin\big(\frac{\theta(z, \rho, t)}{2}\big)
\end{pmatrix},
\end{eqnarray}
where 
\begin{eqnarray}
\label{R}
R(z, \rho, t)=\sqrt{|\tilde{\psi}_a(z, \rho, t)|^2+|\tilde{\psi}_b(z, \rho, t)|^2},
\end{eqnarray}
\begin{eqnarray}
\label{theta}
\theta(z, \rho, t)=2
\;
\text{arctan}\left(\frac{|\tilde{\psi}_b(z, \rho, t)|}{|\tilde{\psi}_{a}(z, \rho, t)|}\right),
\end{eqnarray}
and
\begin{eqnarray}
\label{phi}
\phi(z, \rho, t)=\tilde{\gamma}_b(z, \rho, t)-\tilde{\gamma}_a(z, \rho, t).
\end{eqnarray}
Here, $\tilde{\gamma}_a(z, \rho, t)$ can be interpreted
as an overall spatially dependent phase of the spinor wave function.
This phase has no effect on the physical observables considered 
in this work. 
The quantity
$R(z, \rho, t)$ corresponds to a \textquotedblleft weight''
at each
spatial point.
The spinor dynamics for a given $z$ and $\rho$ is thus conveniently visualized by 
a vector of length $R(z,\rho,t)$ 
on the Bloch sphere. The direction of the vector is given by
$\theta(z,\rho,t)$ and the relative phase $\phi(z,\rho,t)$
 between
the components $ \tilde{\psi}_b(z,\rho,t)$ and 
$\tilde{\psi}_a(z,\rho,t)$.

To visualize the motion of the spinor on the Bloch sphere,
we 
define the local spin expectation values $\sigma_j(z,\rho,t)$,
where $j=x$, $y$, or $z$,
through
\begin{eqnarray}
\label{eq_localspin}
\sigma_j(z,\rho,t)
=
\begin{pmatrix}
[\tilde{\psi}_a(z, \rho, t)]^* \\
[\tilde{\psi}_b(z, \rho, t)]^*
\end{pmatrix}^T
\hat{\sigma}_j
\begin{pmatrix}
\tilde{\psi}_a(z, \rho, t) \\
\tilde{\psi}_b(z, \rho, t)
\end{pmatrix},
\end{eqnarray}
where $\hat{\sigma}_x$, $\hat{\sigma}_y$, and $\hat{\sigma}_z$
denote
the \textquotedblleft usual''
Pauli matrices.
Note that we  use the term
\textquotedblleft spin expectation value'' for convenience
throughout this paper even though our definition in Eq.~(\ref{eq_localspin})
excludes the conventional $\hbar/2$ factor.
Physically, $\sigma_z(z,\rho,t)$
corresponds to the local $(z,\rho)$-specific
population difference at time $t$.
Mathematically, one finds
\begin{eqnarray}
\sigma_x(z,\rho,t)=|R(z,\rho,t)|^2 \cos(\phi(z,\rho,t))\sin(\theta(z,\rho,t)),
\end{eqnarray}
\begin{eqnarray}
\sigma_y(z,\rho,t)=|R(z,\rho,t)|^2 \sin(\phi(z,\rho,t))\sin(\theta(z,\rho,t)),
\end{eqnarray}
and 
\begin{eqnarray}
\sigma_z(z,\rho,t)=|R(z,\rho,t)|^2 \cos(\theta(z,\rho,t)).
\end{eqnarray}

The initial state at $t=0$ 
corresponds to a vector pointing to the north pole on the Bloch sphere. 
From the Hamiltonian $\hat{\tilde{H}}$ [Eq.~\eqref{eqn6}],
it can be seen that the non-vanishing Raman coupling term introduces
a torque along the positive $x$-axis.
Thus, neglecting interactions,
the first $\pi/2$-pulse
rotates the spinor wave function 
by $-\pi/2$
about the $x$-axis on the Bloch sphere.
As a result, the
spinor 
points along the negative $y$-axis on the Bloch sphere 
after the first $\pi/2$-pulse.
Our
numerical mean-field results, which show
that $\theta$ and $\phi$ are approximately equal to $\pi/2$ and 
$-\pi/2$ across the entire BEC after the first $\pi/2$-pulse
are consistent with this simple picture.
Correspondingly, $\sigma_z(z,0,t)$ is approximately zero
and $\sigma_y(z,0,t)$ has---except for a minus
sign---the same $z$-dependence as the 
density
[red dashed and 
black solid lines Fig.~\ref{fig_ramanramsey}(avi)].
We conclude that mean-field effects can,
in a first-order approximation,  be neglected
during the first $\pi/2$-pulse.

During the hold time, 
two effects need to be accounted for:
First, as already pointed out earlier,
the population in state $ \tilde{\psi}_b({\bf{r}},t)$ moves 
relative to that in state $ \tilde{\psi}_a({\bf{r}},t)$.
Second, the phases of the spinor components
$ \tilde{\psi}_a(z,\rho,t)$ 
and $ \tilde{\psi}_b(z,\rho,t)$
evolve independently.
For each $(z,\rho)$,
the combination of these two effects leads to a rotation of the
two-component spinor on the Bloch sphere. 
As an example,
Figs.~\ref{fig_ramanramsey}(avii)-\ref{fig_ramanramsey}(aviii) 
show $\sigma_z(z,0,t)$ and $\sigma_y(z,0,t)$
for two different hold times.
For both hold times, $\sigma_z(z,0,t)$ changes approximately
linearly with $z$ in the region where the two components overlap. This
follows immediately from the approximately parabolic shapes of the
two density components, which are offset from each other:
the difference leads to a term that is, to leading order, linear in $z$.
The local spin expectation value $\sigma_y(z,0,t)$ develops \textquotedblleft wiggles'' 
during the hold time in the
region where the two components overlap. The number of wiggles increases with
increasing hold time. The wiggles arise from the
relative phase dynamics and indicate interference; 
importantly, the
densities do not show any indication of interference 
prior to the application of the second $\pi/2$-pulse.

The second $\pi/2$-pulse \textquotedblleft rotates'' 
$\sigma_y(z,\rho,t)$ to $\sigma_z(z,\rho,t)$.
This can be seen clearly by
comparing Fig.~\ref{fig_ramanramsey}(aix)
with Fig.~\ref{fig_ramanramsey}(avii)
(both these figures are for a hold time of $1$~ms)
and by 
comparing Fig.~\ref{fig_ramanramsey}(ax)
with Fig.~\ref{fig_ramanramsey}(aviii)
(both these figures are for a hold time of $2$~ms).
Since the relative phase information has been \textquotedblleft moved''
from $\sigma_y(z,\rho,t)$ to $\sigma_z(z,\rho,t)$
and since 
$\sigma_z(z,\rho,t)$ is equal
to $| \tilde{\psi}_a(z,\rho,t)|^2-| \tilde{\psi}_b(z,\rho,t)|^2$,
the interference is---after the second $\pi/2$-pulse---visible
in the densities of the components
[Figs.~\ref{fig_ramanramsey}(aiv)-\ref{fig_ramanramsey}(av)].

\subsection{Ramsey-type pulse sequence: Analytical treatment}
\label{sec_theory_ramsey_ana}

The numerical results presented in the previous section
are obtained using the 
scattering lengths $a_{ij}$ for $^{87}$Rb.
Repeating the numerical calculations for equal
scattering lengths $a_{ij}$ reveals that
the effects due to the difference in the scattering lengths
are quite small for the time scales considered in this paper.
Motivated by this observation,
the analytical treatment presented in this section makes the
simplifying assumption that the scattering lengths are all equal
($a_{aa}=a_{ab}=a_{ba}=a_{bb}$).
Moreover, the treatment assumes that 
$\delta_R$ and $t_{\text{start}}$ are equal to zero.
Our analytical model is motivated by
Refs.~\cite{castin1996,other_castin}; however, the application to the Ramsey-type pulse sequence
discussed here has, to the best of our knowledge, not been discussed 
previously.

We assume that the initial state $\tilde{\psi}_a({\bf{r}}, t=0)$ 
is described well within the Thomas-Fermi approximation.
We additionally assume that the population transfer process
commutes with the relative moving and expansion processes 
during the first $\pi/2$-pulse.
Specifically, our analytical model treats
the population transfer
associated with the first $\pi/2$-pulse as occuring instantaneously and then
subsequently treats the
relative moving and expansion of the two components for the duration $\tau_1$
of the actual $\pi/2$-pulse.
For the second $\pi/2$-pulse, we reverse the
order of the operations, i.e., we first treat the 
relative moving and expansion of the two components for the duration $\tau_2$
and then treat the population transfer 
associated with the second $\pi/2$-pulse as occuring instantaneously.
In the following, we provide an analytical framework that
yields approximate 
expressions for $\tilde{\psi}_a(z,\rho,t)$ and $\tilde{\psi}_b(z,\rho,t)$
right after the first instantaneous $\pi/2$-pulse ($t=0^+$), 
during the time $0^+ < t < \tau_1 + t_{\text{hold}} + \tau_2$,
and 
right after the second instantaneous $\pi/2$-pulse ($t=t_{\text{end}}^+$).

Motivated by the discussion
in Sec.~\ref{sec_ramanrabi}, we make the ansatz that 
$|\tilde{\psi}_{a}({\bf{r}},t)|^2$ has 
an inverted parabola-like form during the 
\textquotedblleft effective'' hold time, 
i.e., for $0^+ < t < t_{\text{end}}$,
\begin{eqnarray}
\label{ana_eqn6}
|\tilde{\psi}_{a}({\bf{r}},t)|^2= \nonumber \\
\frac{1}{\lambda_z(t)\lambda_{\rho}^2(t)}
\left[
-\alpha_z\left(\frac{z}{\lambda_z(t)}\right)^2-
\alpha_{\rho}\left(\frac{\rho}{\lambda_{\rho}(t)}\right)^2+
\frac{\mu}{2 g}\right],
\end{eqnarray}
where $\mu$ denotes
the chemical
potential of the initial state,
i.e., of the system
prior to the application of the first $\pi/2$-pulse~\cite{stringariRMP},
\begin{eqnarray}
\mu=
\frac{1}{2} \left[ \omega_\rho^4\omega_z^2 
m^3
\left(\frac{15g}{4\pi}\right)^{2} \right]^{1/5},
\end{eqnarray}
and
\begin{eqnarray}
\label{ana_eqn7}
\alpha_{z/ \rho}=\frac{m\omega_{z/ \rho}^2}{4g}.
\end{eqnarray}
Compared to the \textquotedblleft standard case''~\cite{castin1996},
$\alpha_{\rho}$ and $\alpha_z$ are smaller by a factor of 2
since the population for $t=0^+$ is 
assumed to be equally distributed between the
two components,
i.e., $|\tilde{\psi}_b({\bf{r}},0^+)|=|\tilde{\psi}_a({\bf{r}},0^+)|$.  
In Eq.~(\ref{ana_eqn6}), it is understood that $|\tilde{\psi}_a({\bf{r}},t)|$
is zero when the right hand side of the equation takes negative values.
The dimensionless scaling parameters $\lambda_z(t)$ and $\lambda_{\rho}(t)$
obey the initial conditions
$\lambda_{z}(0^+)=\lambda_{\rho}(0^+)=1$.
The differential equations that govern the time evolution
of  $\lambda_z(t)$
and $\lambda_{\rho}(t)$
are discussed below.
We set $\tilde{\gamma}_a({\bf{r}},0^+)=0$
and assume that the first $\pi/2$-pulse introduces a $-\pi/2$
phase shift onto the second component, i.e., $\tilde{\gamma}_b({\bf{r}},0^+)=-\pi/2$.

For $0^+ < t < t_{\text{end}}$,
$\tilde{\psi}_b({\bf{r}}, t)$ moves with approximately constant velocity $v_z$,
\begin{eqnarray}
v_z = \frac{2\hbar k_R}{m},
\end{eqnarray} 
along the $z$-direction while the center-of-mass 
of $|\tilde{\psi}_a({\bf{r}}, t)|^2$ 
remains essentially unchanged.
Due to the symmetry of the system, we enforce
\begin{eqnarray}
\label{ana_eqn3}
\tilde{\psi}_b({\bf{r}}, t)=\tilde{\psi}_a(v_z t\hat{e}_z-{\bf{r}}, t)\exp \left( -\imath\frac{\pi}{2} \right).
\end{eqnarray}
Thus, once we have expressions for $|\tilde{\psi}_a(v_z t\hat{e}_z-{\bf{r}},t)|$
and $\tilde{\gamma}_a(v_z t\hat{e}_z-{\bf{r}},t)$,
$\tilde{\psi}_b({\bf{r}},t)$ is determined
through Eq.~(\ref{ana_eqn3}).
To eliminate $\tilde{\psi}_b({\bf{r}},t)$,
we insert Eq.~\eqref{ana_eqn3} into the coupled set of
 time-dependent
mean-field equations.
This yields
\begin{eqnarray}
\label{ana_eqn4}
\imath\hbar\frac{\partial}{\partial t}\tilde{\psi}_a({\bf{r}}, t)=
\hat{\tilde{H}}_{\text{hold}}\tilde{\psi}_a({\bf{r}}, t),
\end{eqnarray}
where
\begin{eqnarray}
\label{ana_eqn5}
\hat{\tilde{H}}_{\text{hold}}=
\frac{\hat{\bf{p}}^2}{2m}+g|\tilde{\psi}_{a}({\bf{r}},t)|^2+
g|\tilde{\psi}_{a}(v_z t\hat{e}_z-{\bf{r}},t)|^2.
\end{eqnarray}
From Eq.~\eqref{ana_eqn6}, we find
\begin{eqnarray}
\label{ana_eqn9}
|\tilde{\psi}_{a}(v_z t\hat{e}_z-{\bf{r}},t)|^2 = \nonumber \\
|\tilde{\psi}_a({\bf{r}},t)|^2 +\frac{2\alpha_z v_z tz}
{\lambda_{z}^3(t)\lambda_{\rho}^2(t)}-\frac{\alpha_zv_z^2 t^2}
{\lambda_z^3(t)\lambda^2_{\rho}(t)}.
\end{eqnarray}
Plugging Eq.~\eqref{ana_eqn9} into Eq.~\eqref{ana_eqn5}, we obtain
\begin{eqnarray}
\label{ana_eqn10}
\hat{\tilde{H}}_{\text{hold}}=
\frac{\hat{\bf{p}}^2}{2m}+2g|\tilde{\psi}_{a}({\bf{r}},t)|^2
-F_z(t) z
+C(t).
\end{eqnarray}
Equation~(\ref{ana_eqn10}) 
implies that $C(t)$,
which is independent of ${\bf{r}}$, contributes an overall phase to $\tilde{\psi}_a({\bf{r}},t)$ 
at each time $t$.
The effective time-dependent force $F_z(t)$ along the 
negative $z$-direction,
\begin{eqnarray}
\label{ana_eqn11}
F_z(t)=-\frac{2g\alpha_zv_z t}{\lambda_{z}^3(t)\lambda_{\rho}^2(t)},
\end{eqnarray}  
is due to
the relative motion of the two components with
respect to each other and the mean-field interactions.

\begin{figure}
\vspace*{0cm}
\includegraphics[width=.35\textwidth]{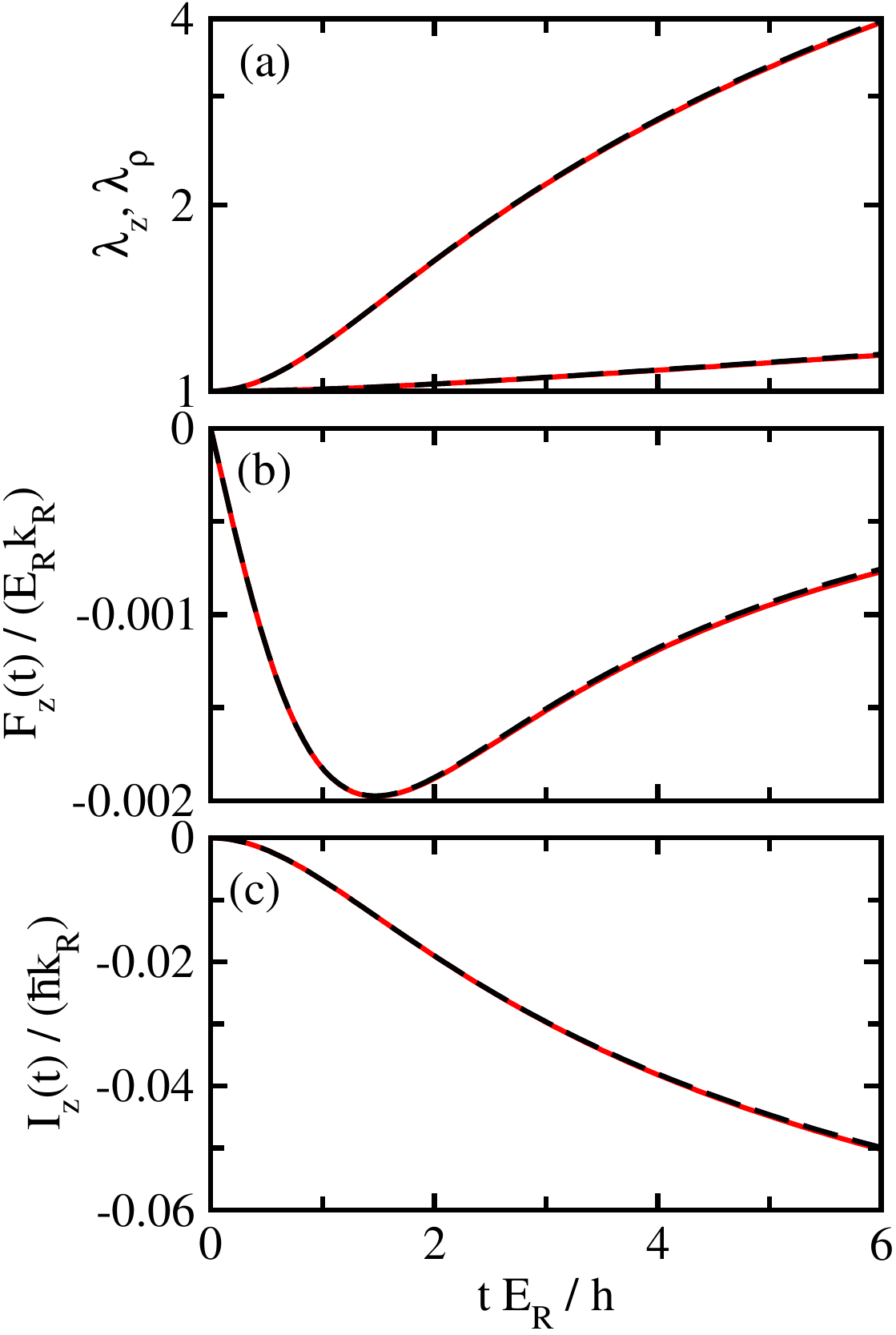}
\caption{Characteristics of the analytical framework
discussed in 
Sec.~\protect\ref{sec_theory_ramsey_ana}:
Scaling parameters, effective force, and impulse
during the hold time of the Raman-Ramsey-type pulse
sequence for a $^{87}$Rb BEC.
Results are shown
for $E_R/h=1960$~Hz, $\Omega_{0,R}=E_R$, 
$\delta_R=0$, $N=3 \times 10^5$,
$\omega_{\rho}= 2 \pi \times 200$~Hz, and 
$\omega_z =  2 \pi \times 40$~Hz.
(a) The red dashed and blue dash-dotted lines
show
the scaling parameters $\lambda_{z}(t)$ and $\lambda_{\rho}(t)$, 
respectively, 
obtained using the formula given in Eqs.~\eqref{lambda_perp} and \eqref{lambda_z};
note that the vertical axis employs a logarithmic 
scale.
(b)
The red dashed line shows the
effective force $F_z(t)$ calculated using our analytical expressions
for the scaling parameters in
Eq.~(\ref{ana_eqn11}).
(c)
The red dashed line shows 
the impulse $I_z(t)$
calculated using our
analytical expressions for the scaling parameters 
in
Eq.~(\ref{ana_eqn12}).
For comparison, the solid lines in panels~(a)-(c)
show results obtained by numerically solving the 
differential equations for $\lambda_{\rho}(t)$
and $\lambda_{z}(t)$.
The excellent agreement between the analytical and numerical results 
validates the use of the analytical expressions for the scaling parameters.
}
\label{ana_fig9}
\end{figure}

We now make the assumption that the effective force term in Eq.~\eqref{ana_eqn10} 
does not
notably affect the time evolution
of the density $|\tilde{\psi}_a({\bf{r}},t)|^2$. Under this assumption,
the time evolution of the scaling factors 
$\lambda_z(t)$ and $\lambda_{\rho}(t)$ 
is governed by
the differential equations derived by Castin and Dum
from the scaling ansatz for
a single-component BEC~\cite{castin1996}:
\begin{eqnarray}
\label{eq_castin1}
\frac{d^2\lambda_z(t)}{dt^2}=\frac{\omega_z^2}{\lambda_\rho^2(t)\lambda_z^2(t)},
\end{eqnarray}
and
\begin{eqnarray}
\label{eq_castin2}
\frac{d^2\lambda_\rho(t)}{dt^2}=\frac{\omega_\rho^2}{\lambda_\rho^3(t)\lambda_z(t)}.
\end{eqnarray}
The black solid lines in Fig.~\ref{ana_fig9}(a)
show $\lambda_{\rho}(t)$ and $\lambda_{z}(t)$,
obtained by solving Eqs.~(\ref{eq_castin1})-(\ref{eq_castin2})
numerically for the same parameters as those employed in Fig.~\ref{fig_ramanramsey}.
Using these solutions, the solid line in Fig.~\ref{ana_fig9}(b)
shows the effective force $F_z(t)$ as a function of the
dimensionless time $t E_R/h$. 
The magnitude of the effective force first increases 
and then decreases with increasing time.
As shown below, 
the turn-around time is,
to leading order, given by the
inverse of the transverse trapping frequency $\omega_{\rho}$.
 
While we assumed that the 
time dynamics 
of $|\tilde{\psi}_a({\bf{r}},t)|$
is largely independent of the effective force $F_z(t)$,
we deduce from Sec.~\ref{sec_ramanramsey_overview} that
the time evolution of the phase $\tilde{\gamma}_a({\bf{r}},t)$
is non-negligibly impacted by $F_z(t)$.
According to the momentum-impulse relationship,
the impulse $I_z(t)$ imparted by the effective force on the system at time $t$ reads 
\begin{eqnarray}
\label{ana_eqn12}
I_z(t)=
\int_{0}^{t}
F_z(\tau)
d \tau .
\end{eqnarray}  
The black solid line in Fig.~\ref{ana_fig9}(c)
shows that
the magnitude of $I_z(t)$ increases monotonically
with increasing effective hold time.
Using that the change of the momentum during the hold
time is equal to
$I_z(t)$, we estimate that
the effective force $F_z(t)$ changes the phase
$\tilde{\gamma}_a({\bf{r}},t)$ 
by
$\phi_{\text{lin}}(z,t)$,
\begin{eqnarray}
\label{eq_philinear}
\phi_{\text{lin}}(z,t) = \frac{I_z(t)}{\hbar} z,
\end{eqnarray}
where $0^+ < t < t_{\text{end}}$.
We refer to this phase as \textquotedblleft linear phase'' since it depends linearly
on $z$.
It vanishes in the limit that
the population of state $\tilde{\psi}_{b}({\bf{r}},t)$ does not move
relative 
to that in
state $\tilde{\psi}_a({\bf{r}},t)$.

The expansion of the two components for $0^+ < t < t_{\text{end}}$ 
introduces an additional contribution to the phase $\tilde{\gamma}_a({\bf{r}},t)$,
which we refer to as
quadratic phase $\phi_{\text{quad}}(z,\rho,t)$ 
due to its quadratic dependence on $z^2$ and $\rho^2$.
The quadratic phase is independent of 
$v_z$ and analogous to the
phase that develops during the expansion of a single-component BEC~\cite{castin1996},
\begin{eqnarray}
\phi_{\text{quad}}(z,\rho,t)=
\frac{m z^2}{2\hbar\lambda_{z}(t)}\frac{d\lambda_z(t)}{dt}+
\frac{m \rho^2}{2\hbar\lambda_{\rho}(t)}\frac{d\lambda_{\rho}(t)}{dt}.
\end{eqnarray}
Combining the linear and quadratic phases, 
the expression for $\tilde{\gamma}_{a}({\bf{r}},t)$ reads
\begin{eqnarray}
\label{eq_gamma_tot}
\tilde{\gamma}_{a}({\bf{r}},t)=
\phi_{\text{lin}}(z,t)
+
\phi_{\text{quad}}(z,\rho,t).
\end{eqnarray}
The second $\pi/2$-pulse, applied at $t=t_{\text{end}}$, 
rotates the spinor at each ${\bf{r}}$ 
by $-\pi/2$ about the $x$-axis on the Bloch sphere.

The division of the phase $\tilde{\gamma}_a({\bf{r}},t)$
into a linear and a quadratic contribution was, to the best of our knowledge, 
first discussed in Ref.~\cite{phillips2000}.
For later work see Refs.~\cite{edwards2011,gupta2011}.
Reference~\cite{phillips2000}
measured,
employing a theory framework motivated by Ref.~\cite{castin1996}, the linear and quadratic phases 
of a $^{23}$Na BEC using matter-wave
Bragg interferometry
(see Ref.~\cite{bongs2001} for a related
measurement of the linear phase). Even though the $\pi/2$ Bragg
pulses are notably shorter than our $\pi/2$-pulses,
the scenario considered in Secs.~\ref{sec_theory_latticeramsey}
and \ref{sec_lattice_ramsey_experiment},
namely, the Ramsey-type pulse sequence for the lattice
coupling case, is closely related to Ref.~\cite{phillips2000}.

To push the analytical treatment even further,
we approximate $\lambda_{z}(t)$ and $\lambda_{\rho}(t)$ 
by~\cite{castin1996}
\begin{eqnarray}
\label{lambda_perp}
\lambda_{\rho}(t)\approx \sqrt{1+\omega_{\rho}^2t^2}
\end{eqnarray}
and
\begin{eqnarray}
\label{lambda_z}
\lambda_{z}(t)\approx \nonumber \\
1+ \left(\frac{\omega_z}{\omega_{\rho}}\right)^2\left[
\omega_{\rho}t \; \arctan(\omega_{\rho}t) -\ln \sqrt{1+\omega_{\rho}^2t^2}\right].
\end{eqnarray}
The red dashed and blue dash-dotted lines in Fig.~\ref{ana_fig9}(a) 
show 
$\lambda_z(t)$ and
$\lambda_{\rho}(t)$ obtained using 
these analytical expressions.
The agreement with the numerical solutions to
Eqs.~(\ref{eq_castin1})-(\ref{eq_castin2}) [see the solid
lines in Fig.~\ref{ana_fig9}(a)] is excellent.
As a consequence, 
the effective force $F_z(t)$ and impulse $I_z(t)$, calculated
using 
the approximate results for the scaling parameters,
nearly coincide with the results that are obtained using the
numerically determined scaling factors
[compare the red dashed and black solid
lines in Figs.~\ref{ana_fig9}(b)
and \ref{ana_fig9}(c)].
While Fig.~\ref{ana_fig9} focuses on one specific parameter combination, 
similarly convincing agreement is found for other parameter combinations.

To understand the non-monotonic behavior of the effective force
displayed in Fig.~\ref{ana_fig9}(b), 
we plug Eqs.~\eqref{lambda_perp} and~\eqref{lambda_z} into Eq.~\eqref{ana_eqn11}.
Taylor-expanding around small $\omega_z/\omega_{\rho}$ and 
neglecting terms of order $(\omega_z/\omega_{\rho})^2$ and higher,
we obtain 
\begin{eqnarray}
\label{force_taylor}
F_{z}(t)
\approx
-\frac{2g\alpha_z v_z}{\omega_{\rho}}
\left(\frac{1}{\omega_{\rho} t}+\omega_{\rho} t\right)^{-1}.
\end{eqnarray}
Thus, 
for $\omega_{\rho} t \ll 1$
and $\omega_{\rho} t \gg 1$, 
$F_z(t)$
is proportional to $-\omega_{\rho} t$
and $-(\omega_{\rho} t)^{-1}$, respectively.
Within the approximations made,
$F_z(t)$ takes on a global minimum for $\omega_{\rho} t=1$.
For comparison, the minimum in Fig.~\ref{ana_fig9}(b)
occurs at $t E_R / h \approx 1.5$, which corresponds to
$\omega_{\rho} t \approx 0.96$.

Equipped with fully analytical expressions for $\tilde{\psi}_a({\bf{r}},t)$
and $\tilde{\psi}_b({\bf{r}},t)$,
we are ready to compare the spin dynamics obtained within
this Thomas-Fermi approximation-like framework to the spin
dynamics obtained by solving the 
Gross-Pitaevskii equation numerically.
To this end, the third and fourth columns of Fig.~\ref{fig_ramanramsey}
show the same observables as the first and second columns.
While the first and second columns are obtained---as discussed
in detail in Sec.~\ref{sec_ramanramsey_overview}---by analyzing
the solutions to the full Gross-Pitaevskii equation,
the third and fourth columns are obtained using our 
fully analytical solutions derived above.
A quick comparison indicates 
that the overall agreement is strikingly good. This 
{\em{a posteriori}} justifies
the assumptions made
in developing the analytical framework presented in this section.
Most importantly, the good agreement allows us to 
unambiguously state that both the linear phase and the quadratic
phase need to be accounted for to obtain
a faithful description of the 
interference fringes.

Section~\ref{sec_theory_latticeramsey}
returns to the theoretical framework developed in this section.
The analytical framework for the Raman Ramsey-type
sequence, which relies heavily on the assumption that all four coupling strengths $g_{ij}$ are 
(approximately) equal to each other,
cannot be applied directly to the lattice Ramsey-type
sequence since the corresponding two-state model
is described by
coupling strengths $g_{aa}=g_{bb}=g_{ab}/2=g_{ba}/2$.
Despite of this, it is argued in Sec.~\ref{sec_theory_latticeramsey}
that the model developed here provides important insights for the lattice coupling case as well.

\section{Lattice coupling case}
\label{sec_theory_lattice}

\subsection{Two-state model}
\label{two_momentum_component}
\begin{figure}
\vspace*{.2in}
\includegraphics[width=.42\textwidth]{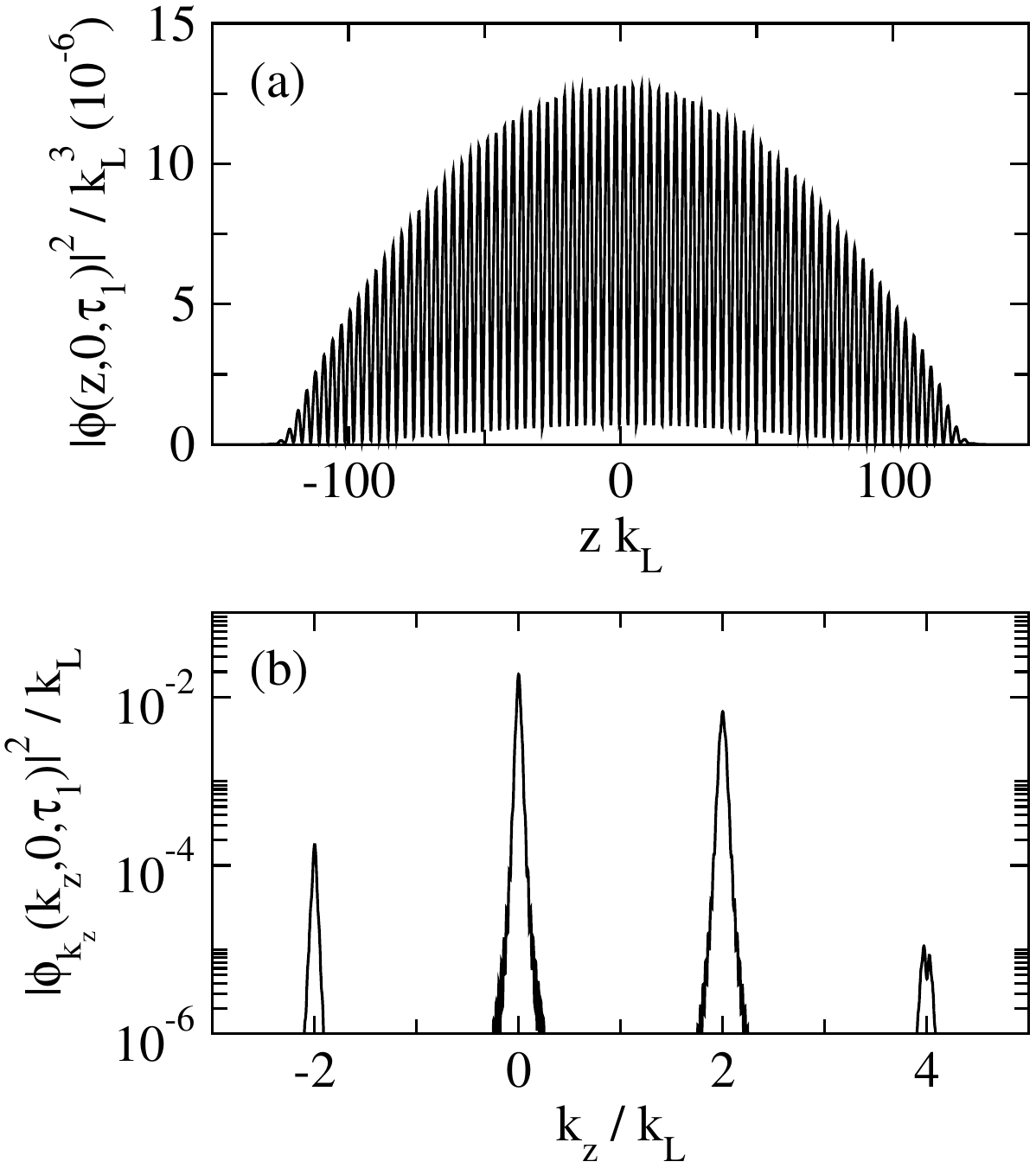}
\caption{$^{87}$Rb BEC density after the application of
a $\pi/2$ lattice pulse 
(numerical results).
The solid lines in (a) and (b) show density cuts as a function of
$z k_L$ (real space) and $k_z/k_L$
(momentum space), respectively, for
$t_{\text{start}}=0$,
$N=3 \times 10^5$, 
$E_L/h=1960$~Hz (corresponding to $k_L=5.81$~$\mu$m$^{-1}$),
$\Omega_{0,L}=E_L$, 
$\delta_L=0$, 
$\omega_{\rho}=2 \pi \times 200$~Hz,
and
$\omega_z=2 \pi \times 40$~Hz.
The real space density cut $|\Phi(z,0,\tau_1)|^2$
is governed by 
fine
oscillations that are related to the fact
that the BEC contains, after the application of the lattice pulse, 
non-zero momentum components.
The momentum space cut 
[$|\Phi_{k_z}(k_z,0,\tau_1)|^2$
is obtained by taking the square of the Fourier transform
of $\Phi(z,0,\tau_1)$] shows that
the BEC density is governed by 
momenta centered around
$\hbar k_z \approx 0$ and
$\hbar k_z \approx 2 \hbar k_L$.}
\label{fig_lattice_overview}
\end{figure}

For the lattice case, all atoms occupy the same hyperfine state; our calculations 
reported below are for
$^{87}$Rb atoms in the 
$\lvert F,m_F \rangle =\lvert  1,-1 \rangle$ state.
Assuming a single mean-field wave function $\Phi({\bf{r}},t)$, the
Hamiltonian $\hat{H}$ to be used in the 
time-dependent Gross-Pitaevskii equation reads~\cite{latticeGP}
\begin{eqnarray} 
\label{eq_gplattice}
\hat{H} = \frac{\hat{\bf{p}}^2}{2m}  + V_{\text{trap}}({\bf{r}},t)+
V_{\text{lat}}({\bf{r}},t) + g_{aa} | \Phi({\bf{r}},t)|^2,
\end{eqnarray}
where the one-dimensional
moving lattice potential $V_{\text{lat}}({\bf{r}},t)$ is given
by
\begin{eqnarray}
\label{eq_gplattice2}
V_{\text{lat}}({\bf{r}},t) = 2 \Omega_L(t) \cos ^2 \left( k_L z - \frac{\omega_L  }{2 } t \right)
\end{eqnarray}
and $\Phi({\bf{r}},t)$ is normalized according to
$\int | \Phi({\bf{r}},t)|^2 d{\bf{r}}=1$.
Figure~\ref{fig_lattice_overview}(a)
shows the density $|\Phi({\bf{r}},t)|^2$, 
obtained by solving the time-dependent
 mean-field 
equation for the
Hamiltonian given in Eq.~(\ref{eq_gplattice}) for typical 
experimental parameters,
as a function of $z$ for $\rho=0$ and a time corresponding
to a $\pi/2$-pulse, i.e., for $t=\pi h/ (2 \Omega_{0,L}$). 
In this example, the BEC is prepared in the ground state 
of the harmonic trap. At time $t=0$,
the trapping potential is turned off and the
lattice with $E_L/h=1960$~Hz, $\omega_L=4 E_L/\hbar$ and $\Omega_{0,L}=E_L$ is flashed on for $0.1276$~ms. 
Here, the coupling strength $\Omega_{0,L}$ is chosen to be
comparable to the chemical potential $\mu$ of the BEC at $t=0$
($\Omega_{0,L} \approx 0.59$~$\mu$).
The size of the BEC does not change notably
during the duration of the lattice pulse: it extends over approximately
$80$ lattice sites. 
Figure~\ref{fig_lattice_overview}(a)
shows that the lattice pulse ``imprints"
fine oscillations along the $z$-direction onto the mean-field density.

To facilitate the analysis, it is desirable to bring out
the intrinsic dynamics by rotating the lattice induced oscillations away.
As we discuss in the next paragraphs,
this can be accomplished within the framework of
an approximate two-state model, which
assumes that the BEC only occupies momenta along the $z$-direction
near $\hbar k_z = 0$ and $\hbar k_z = 2 \hbar k_L$ and not
near $n \hbar k_L$ with $n=-2, \pm 4, \pm 6, \cdots$.
This assumption is well justified for the 
example shown in Fig.~\ref{fig_lattice_overview}(a). The density cut 
in momentum space [Fig.~\ref{fig_lattice_overview}(b)]
shows peaks centered near $\hbar k_z=0$ and
$\hbar k_z=2 \hbar k_L$; the populations of these peaks
are $65.46$~\% and $33.79$~\%,
respectively. Since the peaks centered near $\hbar k_z=-2 \hbar k_L$
and $\hbar k_z=4 \hbar k_L$ have tiny populations ($0.665$~\% and $0.084$~\%, respectively),
the two-state model developed below is expected to capture the dynamics of this
system semi-quantitatively.
More generally, the applicability of the two-state model 
requires that the lattice pulse or pulses are sufficiently short and sufficiently weak.
The two-state model introduced below 
can be improved systematically by accounting for successively more 
\textquotedblleft momentum components'', i.e., by increasing the number of $n$ values included
in Eq.~(\ref{eq_ansatz_twostate}).
In the limit of an infinite state model that accounts for
all $n$ ($n=0, \pm 2,\cdots$)
the description is equivalent to that captured by the
original mean-field Hamiltonian [Eq.~(\ref{eq_gplattice}) with
$V_{\text{lat}}({\bf{r}},t)$ given by Eq.~(\ref{eq_gplattice2})].

To derive the  two-state model,
we make the ansatz~\cite{work_on_two_mode_c_number1,work_on_two_mode_c_number2}
\begin{eqnarray}
\label{eq_ansatz_twostate}
\Phi({\bf{r}},t)= \tilde{\psi}_a({\bf{r}},t) + 
\tilde{\psi}_b({\bf{r}},t) \exp ( 2 \imath  k_L z),
\end{eqnarray}
where $\tilde{\psi}_a({\bf{r}},t)$ and
$\tilde{\psi}_b({\bf{r}},t)$ 
are assumed to be localized in the vicinity
of the momenta $\hbar k_z=0$
and  $\hbar k_z = 2 \hbar k_L$, 
respectively.
The functions $\tilde{\psi}_a({\bf{r}},t)$ and
$\tilde{\psi}_b({\bf{r}},t)$ are normalized according to
Eq.~(\ref{eq_norm1}) and Eq.~(\ref{eq_norm2}) with 
$\psi_j({\bf{r}},t)$ replaced by $\tilde{\psi}_j({\bf{r}},t)$.
Since the widths of the momentum distributions 
associated with the
states
$\tilde{\psi}_a({\bf{r}},t)$ and
$\tilde{\psi}_b({\bf{r}},t)$ 
are assumed to be narrow compared to $2 \hbar k_L$
[this is, indeed, the case for the example shown in Fig.~\ref{fig_lattice_overview}(b)],
we demand that the 
``separation condition" 
\begin{eqnarray}
\label{separation_condition}
\int 
\exp (\imath 2 k_L z )
 \tilde{\psi}_a({\bf{r}},t) \left[\tilde{\psi}_b({\bf{r}},t)\right]^* d{\bf{r}} = 0
\end{eqnarray}
holds.

Following the standard mean-field approach, we write
the $N$-body wave function 
as
a product
over single-particle orbitals,
namely as 
$\Phi({\bf{r}}_1,t)\Phi({\bf{r}}_2,t) \cdots \Phi({\bf{r}}_N,t)$.
Variation of 
the energy functional with respect
to $[\tilde{\psi}_a({\bf{r}},t)]^*$ and $[\tilde{\psi}_b({\bf{r}},t)]^*$
then yields two coupled
non-linear equations, namely
Eq.~(\ref{eqn3}) with $\hat{H}$ replaced by
$\hat{\tilde{H}}_{\text{2-st}}$, where
\begin{widetext}
\begin{eqnarray}
\label{coupled_GP_5}
\hat{\tilde{{H}}}_{\text{2-st}}= 
&&
\left(\frac{\hat{\bf{p}}^2}{2m}+V_{\text{trap}}({\bf{r}}, t)\right)\otimes I_2 
 +
\begin{pmatrix}
  &g|\tilde{\psi}_a({\bf{r}},t)|^2+2g|\tilde{\psi}_b({\bf{r}},t)|^2 & 0\\
  & 0 & 2g|\tilde{\psi}_a({\bf{r}},t)|^2+g|\tilde{\psi}_b({\bf{r}},t)|^2\\
\end{pmatrix} + \\ \nonumber
&& 
\begin{pmatrix}
  & 0 & \frac{\Omega_L(t)}{2}\\
  & \frac{\Omega_L(t)}{2} & \frac{2\hbar k_{L} \hat{p}_z}{m}+\delta_L\\
\end{pmatrix}.
\end{eqnarray}
\end{widetext}
In deriving Eq.~(\ref{coupled_GP_5}),
we assumed that integrals such as
\begin{eqnarray}
\int |\tilde{\psi}_j({\bf{r}},t)|^2 
\exp( \pm n \imath k_L z) d {\bf{r}}
\end{eqnarray}
($n=2, 4,\cdots$),
which have rapidly oscillating integrands, vanish.
This means that portions of the kinetic energy,
lattice potential, trap potential, and mean-field energy 
contributions are neglected in Eq.~(\ref{coupled_GP_5}).

Comparison of the approximate two-state lattice Hamiltonian 
[Eq.~\eqref{coupled_GP_5}] and the rotated Raman Hamiltonian
[Eq.~\eqref{eqn6}]
shows that 
the two Hamiltonians agree
if we enforce that $E_R$, $\Omega_R(t)$, and $\delta_R$
are equal to $E_L$, $\Omega_L(t)$, and $\delta_L$, respectively
and
if additionally the following holds: $g_{aa}=g_{bb}=g$ and $g_{ab}=2g$.
For the $F=1$ hyperfine manifold of $^{87}$Rb, the
mean-field interactions of the two Hamiltonians do not agree
since we have
$g_{aa} \approx g_{bb} \approx g_{ab}$. Consequently, the
dynamics for the Raman coupled and lattice coupled systems are expected to differ
even if the single-particle coupling mechanisms are characterized by 
matching parameters. 
In what follows, we will focus on the interaction-induced differences.

Since 
the population in state $\tilde{\psi}_a({\bf{r}},t)$ 
[$\tilde{\psi}_b({\bf{r}},t)$] experiences a mean-field interaction
due to the population in state $\tilde{\psi}_b({\bf{r}},t)$ 
[$\tilde{\psi}_a({\bf{r}},t)$] that 
is about two times larger
for the lattice coupled Hamiltonian than for the Raman coupled Hamiltonian,
the lattice coupled system has a 
stronger tendency to phase separate than the Raman coupled system
(this argument uses the fact that $g$ is positive for the $F=1$ hyperfine manifold of $^{87}$Rb).
Phase separation has been discussed 
in the literature in the context of multi-component
BECs~\cite{phase_separation}.
The framework developed here may also provide
an intuitive understanding of the formation of the ferromagnetic domains
observed in Ref.~\cite{chin2013}.

The difference between the Raman and lattice coupling cases can
also be interpreted from an alternative viewpoint.
To this end, we 
rewrite the mean-field terms from Eq.~(\ref{coupled_GP_5})
as 
\begin{eqnarray}
\label{eq_mf_effective1}
g|\tilde{\psi}_a({\bf{r}},t)|^2+2g|\tilde{\psi}_b({\bf{r}},t)|^2= \nonumber \\
 g_{\text{eff}} [ |\tilde{\psi}_a({\bf{r}},t)|^2+|\tilde{\psi}_b({\bf{r}},t)|^2]
- g|\tilde{\psi}_a({\bf{r}},t)|^2
\end{eqnarray}
and
\begin{eqnarray}
\label{eq_mf_effective2}
g|\tilde{\psi}_b({\bf{r}},t)|^2+2g|\tilde{\psi}_a({\bf{r}},t)|^2= \nonumber \\
 g_{\text{eff}} [ |\tilde{\psi}_a({\bf{r}},t)|^2+|\tilde{\psi}_b({\bf{r}},t)|^2]
- g|\tilde{\psi}_b({\bf{r}},t)|^2,
\end{eqnarray}
where $g_{\text{eff}}$ is defined to be equal to $2g$.
The right hand sides of Eqs.~(\ref{eq_mf_effective1}) and (\ref{eq_mf_effective2}) suggest that the 
difference between the lattice and Raman coupling cases is due to two things:
First, $g_{aa}$, $g_{bb}$, and $g_{ab}$ can be identified to be equal to $g_{\text{eff}}$, suggesting 
that the lattice coupled system is characterized by a two times
stronger 
repulsion than the Raman coupled system.
Second, there exists an effective on-site attraction in the two-state model of the lattice coupled system
of strength
$-g$~\cite{gadway_lattice}  that has no analog in the Raman coupled system.

We emphasize that
the two interpretations introduced above
are consistent with the scattering diagram arguments
outlined in Sec.~\ref{sec_introduction}.
The ``factor of 2'' in the second $2 \times 2$ matrix on the right hand side of
Eq.~(\ref{coupled_GP_5}) is due to non-vanishing scattering matrix elements;
the analogous scattering matrix elements vanish in the Raman coupled case due to the orthogonality
of the two different hyperfine states.

\subsection{Rabi oscillations: Theory overview}
\label{sec_theory_latticerabi}

\begin{widetext}

\begin{figure}
\includegraphics[width=.7\textwidth]{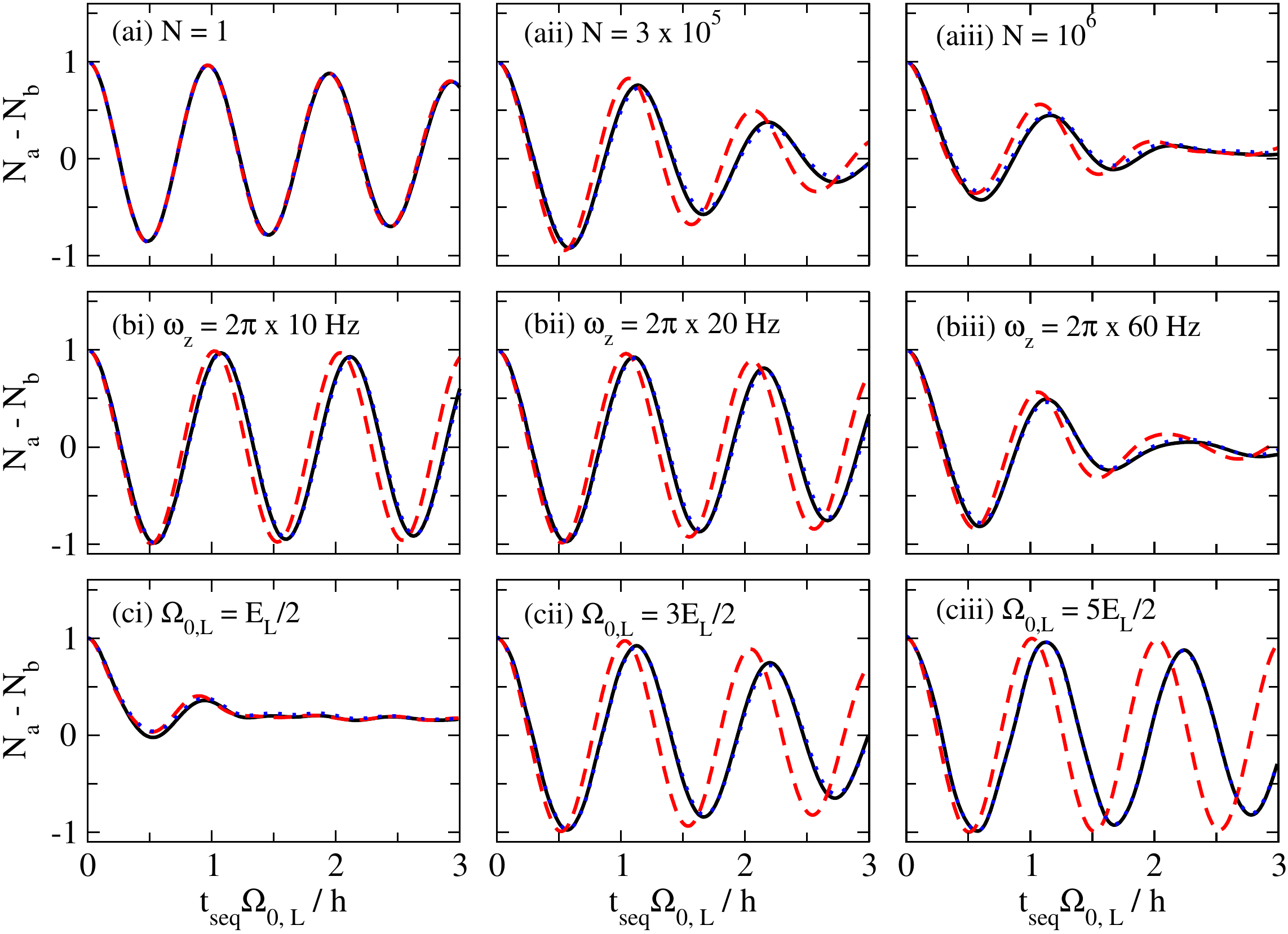}
\caption{Rabi oscillations for lattice coupling case
 (numerical results).
The 
lines show the difference $N_a-N_b$
between the fractional populations
as a function of the dimensionless time 
$t_{\text{seq}} \Omega_{0,{{L}}}/h$
for
$t_{\text{start}}=0$, $E_L/h=1960$~Hz, $\delta_L=0$,
 and $\omega_{\rho}=2 \pi \times 200$~Hz.
The black solid, red dashed, and blue dotted lines show results obtained by solving the time-dependent
mean-field equation 
for the full lattice Hamiltonian 
[Eqs.~(\ref{eq_gplattice}) and (\ref{eq_gplattice2})],
the approximate two-state Hamiltonian [Eq.~(\ref{coupled_GP_5})], and the approximate four-state Hamiltonian 
(this Hamiltonian is not written out explicitly in the text).
The black solid and blue dotted lines nearly coincide
(in particular, the blue dotted lines are hardly visible on the scale shown).
(ai)-(aiii) Changing the particle number $N$ 
(the values are given in the panels).
The weak angular trapping frequency is
$\omega_z=2 \pi \times 40$~Hz
and the coupling strength is $\Omega_{0,L}=E_{{L}}$. 
(bi)-(biii) Changing the angular trapping frequency $\omega_z$
(the values are given in the panels).
The coupling strength is $\Omega_{0,L}=E_{{L}}$ and the number
of particles is $N=3 \times 10^5$. 
(ci)-(ciii) Changing the coupling strength $\Omega_{0,{L}}$
(the values are given in the panels).
The number of particles is $N=3 \times 10^5$ and the weak trapping
frequency is $\omega_z= 2 \pi \times 40$~Hz.
}
\label{fig_lattice_rabi}
\end{figure}    

\end{widetext}

This section discusses lattice coupling induced Rabi oscillations.
Figure~\ref{fig_lattice_rabi}
compares the mean-field results for the full lattice Hamiltonian
[Eqs.~(\ref{eq_gplattice}) and (\ref{eq_gplattice2}); solid 
black lines]
with those obtained using the approximate two- and four-state
Hamiltonians (red dashed and blue dotted lines, respectively).
For all 9 parameter combinations considered in
Fig.~\ref{fig_lattice_rabi}, the four-state model reproduces the 
dynamics obtained for the full mean-field lattice Hamiltonian extremely well.
While the two-state model results deviate somewhat from the 
results for the full lattice Hamiltonian, 
the two-state model captures the main features 
of the Rabi oscillations such as the 
change of the damping of the Rabi oscillations
with increasing number of particles
[Figs.~\ref{fig_lattice_rabi}(ai)-(aiii)] and with increasing
strength of the weak trapping frequency [Figs.~\ref{fig_lattice_rabi}(bi)-(biii)].
Moreover, the rapid reduction of the oscillation amplitude for small coupling
strength [Fig.~\ref{fig_lattice_rabi}(ci) is for $\Omega_{0,L}= E_L/2$]
is remarkably well captured by the approximate two-state model.
For larger lattice strengths [Figs.~\ref{fig_lattice_rabi}(bii) and (biii)
are for $\Omega_{0,L}=3 E_L/2$ and $5 E_L/2$, respectively],
in contrast,
the two-state model captures 
the period of the Rabi oscillations
comparatively poorly.
The reason is that larger lattice coupling
strengths  
lead to enhanced and non-negligible occupations of momenta centered near
$\hbar k_z \approx -2 \hbar  k_L$ and $\hbar k_z \approx 4 \hbar k_L$.

Since the approximate two-state model provides a 
qualitatively and for some parameter combinations even a (semi-)quantitatively correct
description of the dynamics, it is instructive to compare the 
Rabi oscillations for the lattice and Raman coupled systems. 
If the two-state lattice model is exact, the difference between the Rabi oscillations for the 
lattice and Raman coupled systems will be---assuming that the 
small differences between $g_{aa}$, $g_{bb}$, and $g_{ab}$ 
do not play a role---solely due to the \textquotedblleft factor of 2'' discussed in Sec.~\ref{two_momentum_component}.
The parameters in Figs.~\ref{fig_ramanrabi} and \ref{fig_lattice_rabi} are chosen such that 
the solid line
in Fig.~\ref{fig_ramanrabi}(a) can be directly compared with the curves in Fig.~\ref{fig_lattice_rabi}(ai),
the dashed line
in Fig.~\ref{fig_ramanrabi}(a) 
with the curves in Fig.~\ref{fig_lattice_rabi}(aii),
and the dotted line
in Fig.~\ref{fig_ramanrabi}(a) 
with the curves in Fig.~\ref{fig_lattice_rabi}(aiii).
An analogous correspondence exists for Fig.~\ref{fig_ramanrabi}(b) and Figs.~\ref{fig_lattice_rabi}(bi)-(biii)
as well as 
for Fig.~\ref{fig_ramanrabi}(c) and Figs.~\ref{fig_lattice_rabi}(ci)-(ciii).
A careful comparison of Figs.~\ref{fig_lattice_rabi}
and \ref{fig_ramanrabi}
indicates that the most prominent effect of the factor of 2 
is to significantly 
enhance the damping or
dephasing of the Rabi oscillations.  

In what follows we attempt to pinpoint why the factor of 2 (lattice
coupling case)
enhances the damping compared to the case where this factor is equal to 1
(Raman coupling case).
To start this discussion, we remind the reader that the analytical treatment in 
Sec.~\ref{sec_theory_ramsey_ana}, which assumes
vanishing detuning, relies heavily on the assumption that there exists
a symmetry between the components
$\tilde{\psi}_a({\bf{r}},t)$ and
 $\tilde{\psi}_b({\bf{r}},t)$
[see Eq.~(\ref{ana_eqn3})]. 
In fact,
one can show that this symmetry
is---within the Thomas-Fermi approximation---an exact symmetry
provided $g_{aa}=g_{bb}=g_{ab}$.
Intuitively, this symmetry can be understood by realizing that the 
strength of the scattering between
two atoms in the same hyperfine state is identical to that 
of the scattering between two atoms in
different hyperfine states. This implies that neither the 
two-body interactions nor the 
Raman coupling (recall, we are considering the zero detuning
scenario)
bias populations to one hyperfine state over another.
In the lattice coupling case with $\delta_L=0$, the factor of 2 breaks the symmetry.
The effective attractive on-site interactions
[see the discussion in the context of Eqs.~(\ref{eq_mf_effective1})
and (\ref{eq_mf_effective2})], which can alternatively be 
interpreted as effective repulsive off-site interactions,
favor configurations that reduce the overlap between the densities
$|\tilde{\psi}_a({\bf{r}},t)|^2$ and
 $|\tilde{\psi}_b({\bf{r}},t)|^2$. 
 Since the effective repulsive off-site interactions
 depend on the density, they
 vary spatially. This spatial dependence
 can result in
 a shape of the density 
 $|\tilde{\psi}_a({\bf{r}},t)|^2$
 that is different from that of the density 
 $|\tilde{\psi}_b({\bf{r}},t)|^2$.
 If this occurs, 
 the fractional population difference varies locally, leading 
 to a spatially dependent population transfer 
 and, correspondingly, a damping or dephasing of the
 Rabi oscillations. In a complementary picture, the
 effective repulsive off-site interactions can be thought of
 as an effective spatially and temporally varying
 coupling term. In this picture, the damping of the Rabi oscillations emerges naturally.
 Section~\ref{sec_theory_latticeramsey} makes this  
 discussion concrete for
a $\pi/2$ pulse (first step of the Ramsey-type sequence).

\subsection{Rabi oscillations: Theory-experiment comparison}
\label{sec_compare_latticeramsey}

\begin{figure}[t]
\includegraphics[width=0.4\textwidth]{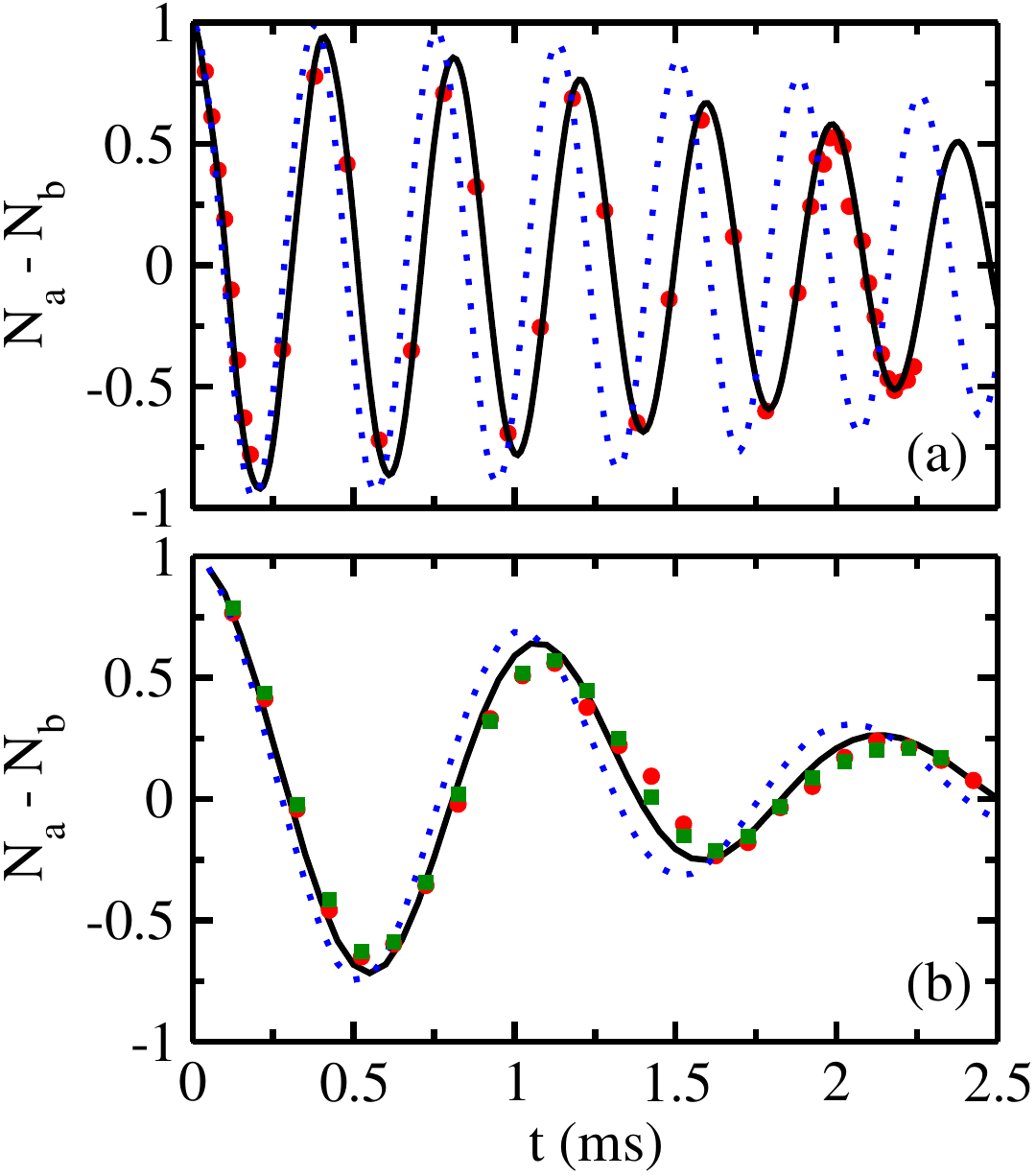}
\caption{Theory-experiment comparison for lattice Rabi oscillations
with ``strong" and ``weak" coupling strengths
for a $^{87}$Rb BEC.
The symbols show experimental data and the black solid lines show
results from the Gross-Pitaevskii simulations for the full lattice Hamiltonian.
For comparison, the blue dotted lines show results obtained 
for the approximate
two-state model.
The experimental parameters common to both panels are
$E_L/h=1080$~Hz
and
$t_{\text{start}}=0.5$~ms.
(a) ``Strong coupling" ($\Omega_{0,L}/h=2646$~Hz): The experimentally measured parameters are
$\omega_x = 2 \pi \times 172$~Hz,
$\omega_y = 2 \pi \times 139$~Hz,
$\omega_z = 2 \pi \times 33.6$~Hz,
$\delta_L/h=-264$~Hz, and
$N=1.1 \times 10^5$.
The calculations set $\omega_{\rho}$ equal to the mean of $\omega_x$ and $\omega_y$;
all other parameters are taken from the experiment.
 The chemical potential $\mu$ prior to turning off the trap is
 $1.439$~$E_L$.
 The mean-field energy per particle  prior to turning off the trap and after the
 $0.5$~ms expansion is
 $0.8881$~$E_L$ and $0.7149$~$E_L$,
 respectively.
 The red circles show the result from one experimental run.
(b) ``Weak coupling" ($\Omega_{0,L}/h=980$~Hz): 
The experimentally measured parameters are
$\omega_{\rho} = 2 \pi \times 146$~Hz,
$\omega_z = 2 \pi \times 28$~Hz,
$\delta_L=0$, and
$N=2.7 \times 10^5$. 
The transverse trap frequency is determined by
performing measurements along
one axis.
The calculations
use the parameters from the experiment.
The chemical potential $\mu$ prior to turning off the trap is
 $1.830$~$E_L$.
 The mean-field energy per particle  prior to turning off the trap and after the
 $0.5$~ms expansion is
 $1.129$~$E_L$ and $0.9297$~$E_L$,
 respectively. 
 The red circles and green squares show the results from two
 separate experimental runs.
}
\label{fig_latticerabiexp}
\end{figure}

The symbols in Figs.~\ref{fig_latticerabiexp}(a)
and
\ref{fig_latticerabiexp}(b)
 show 
experimental data for Rabi oscillations induced by a moving
optical lattice with weak and 
strong coupling, respectively.
The excellent agreement between the
 solutions to the Gross-Pitaevskii equations
 for the full
 lattice Hamiltonian (solid lines)
 and the experimental data indicates that the 
 experiments operate in the mean-field regime, i.e.,
 the Gross-Pitaevskii framework captures the
 population transfer between the two momentum components
 quantitatively.
 The approximate two-state model (dotted lines) provides, as already discussed in the 
 previous section, a semi-quantitative
 description of the lattice-induced Rabi oscillations
 in the weak coupling regime [Fig.~\ref{fig_latticerabiexp}(b)];
 as such, it provides
 a meaningful conceptual framework for interpreting the results and contrasting the 
 lattice- and Raman-induced Rabi oscillations.

A fit of the Rabi oscillation data obtained by solving the Gross-Pitaevskii equation for the full lattice Hamiltonian to Eq.~(\ref{eq_fit}) yields coupling strengths that are, respectively, 5~\% and 6~\% lower than those used 
in the simulations.
This shows that the interactions do impact 
the Rabi oscillations and that calibration of the experimental lattice strength
needs to proceed with care.
We note that the fit
to the data in Fig.~\ref{fig_latticerabiexp}(a)
has a significantly lower $\chi^2$ than the fit to the
data
in Fig.~\ref{fig_latticerabiexp}(b).
For the experimental data shown in Fig.~\ref{fig_latticerabiexp},
the coupling strength is calibrated by inducing 
Rabi oscillations of a 
very dilute $^{87}$Rb BEC for a relative large lattice coupling strength;
in this case, the Rabi oscillation data display essentially no
damping. This calibration run
yields a power-to-coupling-strength conversion.
Assuming that the coupling strength scales
as the square root of the power, 
the calibration curve can be used in subsequent 
science runs
that operate at other powers. The outlined approach assumes
that the power fluctuations are negligible over the course of 
several hours; we have checked that this is the case in our setup.

\subsection{Ramsey-type pulse sequence: Numerical results}
\label{sec_theory_latticeramsey}

\begin{widetext}

\begin{figure}[t]
\includegraphics[width=1\textwidth]{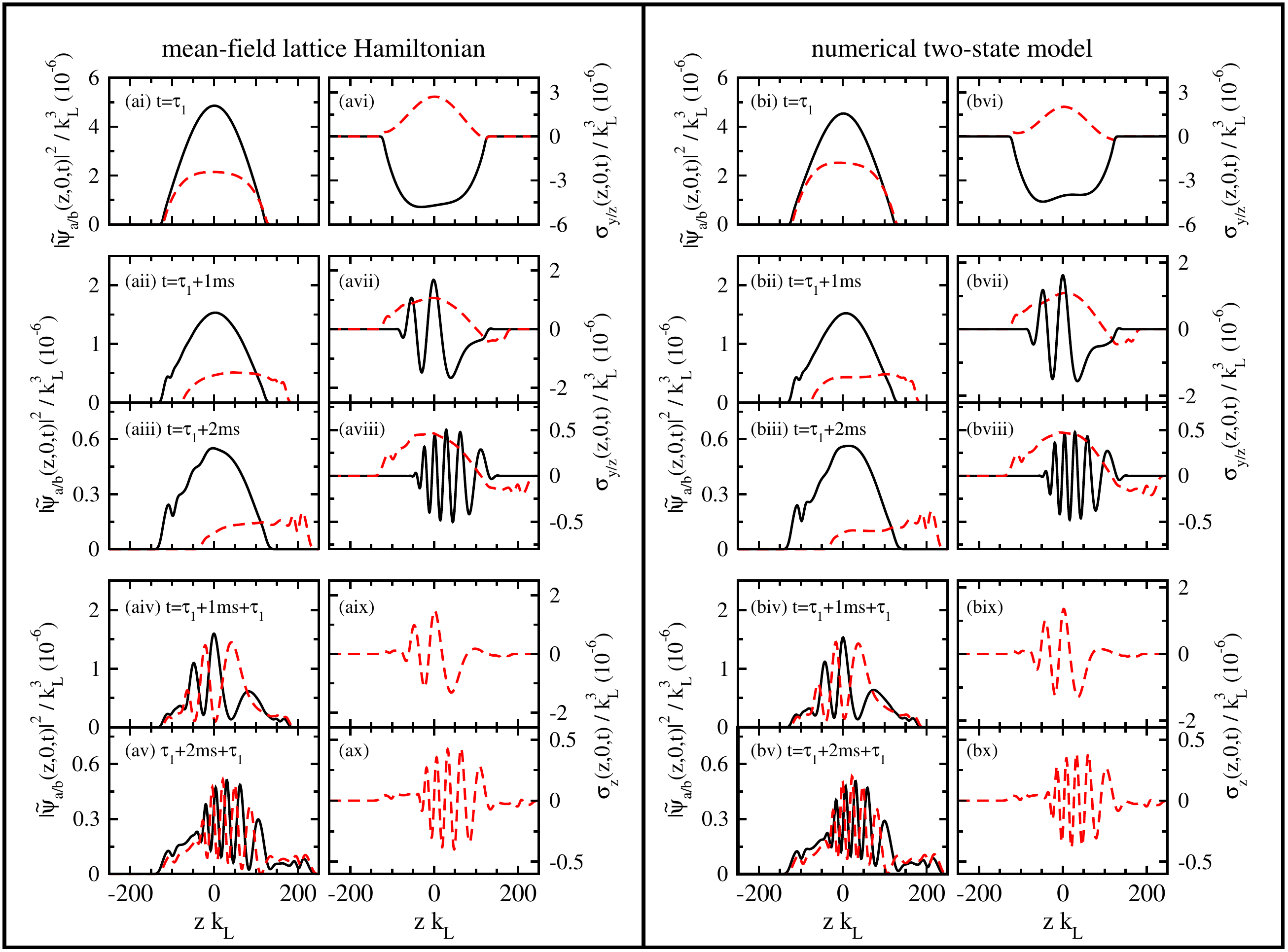}
\caption{Density cuts and local
spin expectation values
for the Ramsey-type pulse sequence 
with $E_L/h=1960$~Hz, $\Omega_{0,L}=E_L$,
and $\delta_L=0$
(numerical results); these are the same
parameters as used in Figs.~\ref{fig_lattice_overview} and \ref{fig_lattice_rabi}(aii).
The $^{87}$Rb BEC consists of $N=3 \times 10^5$ atoms 
and is prepared in an
axially symmetric trap with $\omega_{\rho} =2 \pi \times 200$~Hz
and $\omega_z= 2 \pi \times 40$~Hz
(these are the same parameters as those used in Fig.~\ref{fig_ramanramsey}).
All results are obtained for $t_{\text{start}}=0$.
The first and second columns are obtained by solving 
the time-dependent mean-field equation
for the full lattice Hamiltonian [Eq.~(\ref{eq_gplattice}) with $V_{\text{lat}}({\bf{r}},t)$
given by Eq.~(\ref{eq_gplattice2})] numerically.
The third and fourth columns show the same observables as 
the first and second columns but are, instead, obtained using the 
approximate two-state model introduced in Sec.~\ref{two_momentum_component};
the agreement is quite good.
The black solid and red dashed lines 
in panels~(ai)-(av)
show the
density profiles $|\tilde{\psi}_a(z,0,t)|^2$ and
$|\tilde{\psi}_b(z,0,t)|^2$, respectively. 
The black solid and red dashed lines 
in panels~(avi)-(ax)
show the
local spin expectation values
$\sigma_y(z,0,t)$ and
$\sigma_z(z,0,t)$,
respectively. 
The time increases from the first row,
to the second/third row, to the fourth/fifth row (the value of the time is given in the panels);
the second and fourth row correspond to a hold time of $1$~ms, and
the third and fifth row correspond to a hold time of $2$~ms.
Unlike in
Fig.~\protect\ref{fig_ramanramsey},
the first $\pi/2$-pulse does not lead to 50/50 mixture.
This population imbalance after the first $\pi/2$-pulse contributes to the development of
unequally spaced interference fringes.
}
\label{fig_lattice_ramsey}
\end{figure}

\end{widetext}

This section discusses numerical results for the 
Ramsey-type pulse sequence with lattice coupling
(vanishing detuning, i.e., $\delta_L=0$). 
Figure~\ref{fig_lattice_ramsey} 
shows lattice coupling
results
for the same parameters as 
used in Fig.~\ref{fig_ramanramsey}
(recall, Fig.~\ref{fig_ramanramsey}
shows results for the Ramsey-type sequence with Raman coupling).
The first and second columns in Fig.~\ref{fig_lattice_ramsey}
are obtained by solving the time-dependent
mean-field equation for the full lattice Hamiltonian 
while the third and fourth columns are
obtained by solving the time-dependent
mean-field equation for the approximate
two-state lattice model.
It can be seen that the results for the approximate two-state lattice Hamiltonian
agree with those for the full lattice Hamiltonian rather well.
Since the approximate two-state lattice model
describes the dynamics faithfully, we use it below to gain insights into the results
after the first $\pi/2$-pulse
[Figs.~\ref{fig_lattice_ramsey}(bi) and \ref{fig_lattice_ramsey}(bvi)], 
after the hold time
[Figs.~\ref{fig_lattice_ramsey}(bii), \ref{fig_lattice_ramsey}(bvii),
\ref{fig_lattice_ramsey}(biii), and \ref{fig_lattice_ramsey}(bviii)], and 
after the second $\pi/2$-pulse
[Figs.~\ref{fig_lattice_ramsey}(biv), \ref{fig_lattice_ramsey}(bix),
\ref{fig_lattice_ramsey}(bv), and \ref{fig_lattice_ramsey}(bx)]. 

After the first $\pi/2$-pulse [Figs.~\ref{fig_lattice_ramsey}(ai) and
\ref{fig_lattice_ramsey}(avi)], the population is
distributed unequally among the two components, i.e., component
$\tilde{\psi}_a({\bf{r}},t)$ has a larger population than component
$\tilde{\psi}_b({\bf{r}},t)$;
a 50/50 mixture is realized for a pulse of length $0.2958$~ms,
i.e., for a pulse that is $2.3$ times longer than the $\pi/2$-pulse
employed in Fig.~\ref{fig_lattice_ramsey}.
 Near the edge of the cloud, the density cuts
$|\tilde{\psi}_a(z,0,\tau_1)|^2$ and $|\tilde{\psi}_b(z,0,\tau_1)|^2$ coincide to a good approximation.
In the central region, in contrast, they differ. While the first component profile, $|\tilde{\psi}_a(z,0,\tau_1)|^2$, approximately follows a Thomas-Fermi profile, the second component profile, $|\tilde{\psi}_b(z,0,\tau_1)|^2$, is flatter than a Thomas-Fermi profile.
Correspondingly,
the local spin expectation value $\sigma_z(z,0,\tau_1)$ has 
a roughly Gaussian shape
as opposed to following a linear curve as in the Raman coupled case.
This indicates,
in agreement with
the more general discussion at the end of Sec.~\ref{sec_theory_latticerabi}, that
population from the center of the cloud is pushed toward the edge of the cloud
due to the larger local effective repulsive off-site interaction at the center of the cloud
compared to the edge. 
Using the Bloch-sphere picture, the spatially and temporally dependent
effective repulsive off-site interaction or
coupling leads to a spatially and temporally dependent torque along the $x$-direction
during the first $\pi/2$-pulse.
This is confirmed by 
the 
local spin expectation value $\sigma_y(z,0,\tau_1)$, whose spatial dependence differs 
from that of the densities of the components. 

Altogether, the discussion  
shows that the interactions can,
for the relatively weak lattice coupling strength of $\Omega_{0,L}=E_{L}$ considered 
in Fig.~\ref{fig_lattice_ramsey},
not be neglected during the first $\pi/2$-pulse, i.e., the lattice coupling is not sufficiently strong to
prevent the system from rearranging structurally.
As a consequence, the Rabi oscillations are damped or 
dephase.
For the parameters chosen in Fig.~\ref{fig_lattice_ramsey}, 
the factor of 2 leads to a notable cloud deformation
during the first $\pi/2$ pulse. For other parameter combinations, the
densities of the components may deform more slowly, 
thereby
leading to a slower damping or 
dephasing 
of the Rabi oscillations.

During the hold time,
the amplitude and phase evolution is, 
as in the Raman coupled case, governed by the 
interplay between the interactions and the expansion.
Because of the deviation of the component densities from the Thomas-Fermi profile and the
``absence of symmetry" (see the discussion above), we 
were not able to develop an analytical framework that describes the dynamics during the hold time.
However, comparing the second and third rows
of Fig.~\ref{fig_lattice_ramsey} with the second and third
rows of Fig.~\ref{fig_ramanramsey},
rough similarities between the time dynamics during the
hold time for the two distinct coupling mechanisms can be recognized.
Thus, while we do not have an analytical description,
the formulation developed in the context
of the Raman coupling case can serve as a crude
zeroth-order guide. 

In what follows, we point out
three aspects that are distinct for the lattice 
coupling case:
(i) The local spin expectation value $\sigma_y(z,0,\tau_1)$
is not symmetric with respect to $z=0$; this asymmetry persists during the hold time.
(ii) During the hold time, the shapes of the densities $|\tilde{\psi}_a(z,0,t)|^2$
and
$|\tilde{\psi}_b(z,0,t)|^2$ continue to change appreciably. 
(iii)
The densities of the components
and the local spin expectation value $\sigma_y(z,0,t)$ develop spatial modulations 
during the hold time in the region where the two components are not
spatially overlapping. These spatial modulations are more
pronounced than in the Raman coupling case.

The second $\pi/2$-pulse \textquotedblleft transfers'' the information encoded in $\sigma_y({\bf{r}},t)$
to the population difference $\sigma_z({\bf{r}},t)$.
Since the cloud expands a fair bit during the hold
time, the interactions can, to a good approximation, be neglected
during the second $\pi/2$-pulse.
Consequently, the resulting densities of the components display
a fringe or interference pattern. However, unlike in the Raman coupled case, the
densities in the lattice coupled case are highly non-symmetric.
As argued above, this asymmetry can be interpreted as a 
fingerprint of the fact that one of the terms on the diagonals in the second
$2 \times 2$ matrix on the right hand side of Eq.~(\ref{coupled_GP_5}) is multiplied 
by a
factor of 2.

\subsection{Ramsey-type pulse sequence:
Theory-experiment comparison}
\label{sec_lattice_ramsey_experiment}

\begin{figure}[t]
\includegraphics[width=0.4\textwidth]{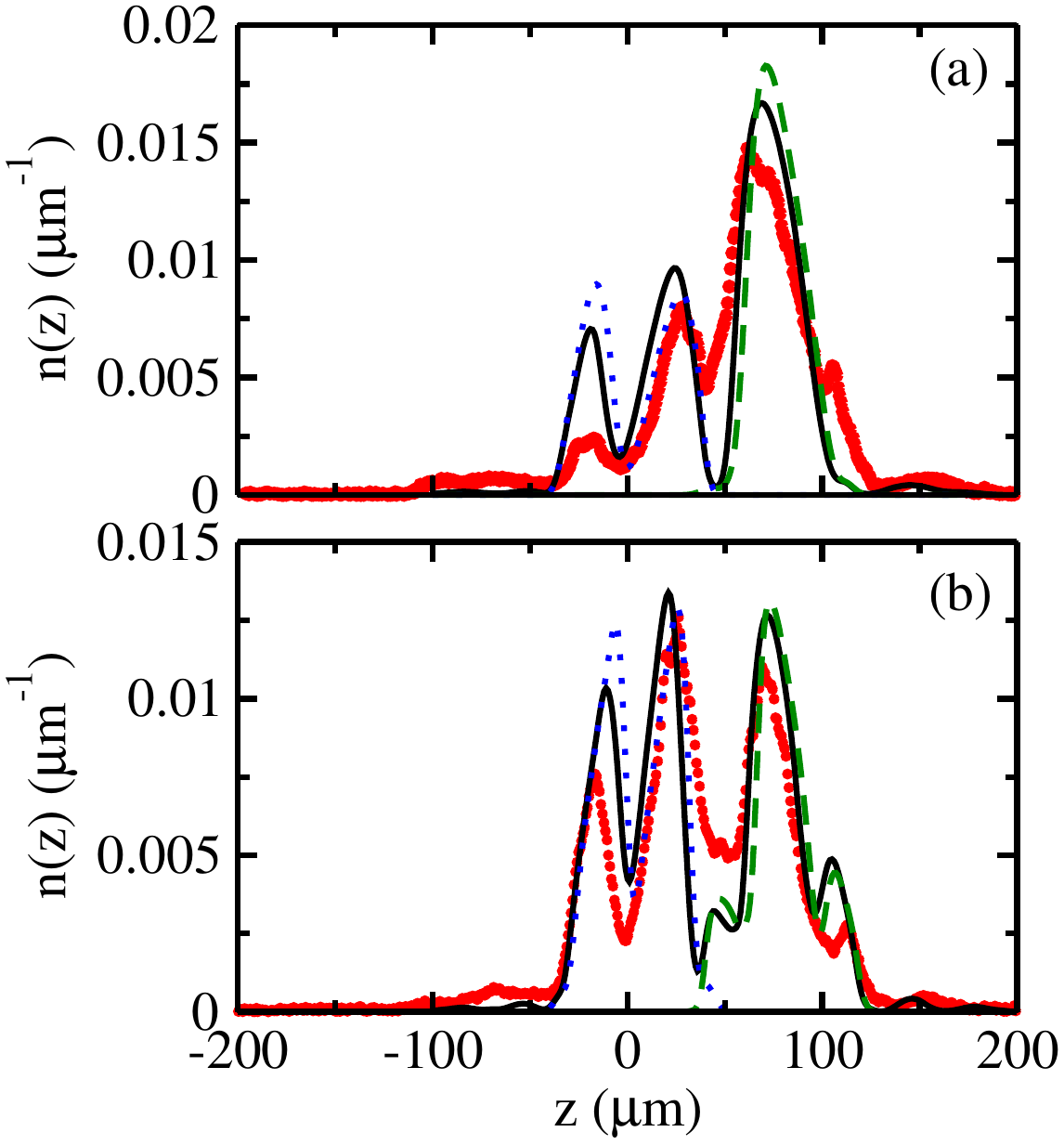}
\caption{Theory-experiment comparison for lattice Ramsey-type pulse sequence
for a $^{87}$Rb BEC  for two different hold times 
$t_{\text{hold}}$
and particle numbers $N$:
(a) 
$t_{\text{hold}} = 0.5$~ms
($N=3.8 \times 10^5$)
and
(b) 
$t_{\text{hold}}=1$~ms
($N=4.3 \times 10^5$).
The red symbols show the experimentally measured 
integrated density $n(z)$,
 $n(z) = \int |\Phi({\bf{r}}, t)|^2 dx dy$, as a function of $z$ 
 for $t_{\text{start}}=0.5$~ms, $t_{\text{ToF}}=12$~ms, $\delta_L=0$,
 and $E_L/h=1080$~Hz; the results shown are from a single experimental run.
The solid black 
lines show results obtained by solving the Gross-Pitaevskii equation
 for the full lattice Hamiltonian for $\Omega_{0,L}/h=1372$~Hz.
 For comparison, the blue dotted and green dashed lines
 show results obtained from using the two-state model (see text). Both sets of
 theory data are convolved using 
a Gaussian with the experimentally measured resolution
width of $2$~$\mu$m.
 The pulse sequence is
 $\tau_1=0.207$~ms (first $\pi/2$-pulse),
 hold for time 
  $t_{\text{hold}}$
 (see above), and 
  $\tau_2 = 0.200$~ms
 (second $\pi/2$-pulse).
  The experimentally determined trap frequencies are  
 $\omega_x =  2 \pi \times 119$~Hz,
 $\omega_y =  2 \pi \times 163$~Hz, and
 $\omega_z =  2 \pi \times 25.7$~Hz;
 the theory calculations set $\omega_{\rho}$
 equal to the mean of $\omega_x$ and $\omega_y$.
% The experimental data are from 180828 data.
}
\label{fig_ramseylatticeexp}
\end{figure}

Figures~\ref{fig_ramseylatticeexp}(a) and 
\ref{fig_ramseylatticeexp}(b) 
compare experimentally determined integrated 
densities (red circles) and  theoretical results
(black solid lines) for the Ramsey-type pulse sequence
with lattice coupling for hold times of $0.5$~ms
and $1$~ms, respectively.
The black solid lines are obtained by solving the 
Gross-Pitaevskii equation
for the full lattice Hamiltonian and explicitly simulating the 
$12$~ms of time-of-flight expansion after the Ramsey-type pulse sequence.
For comparison, the blue dotted and green dashed lines show the
results from the two-state model; the agreement with the full Gross-Pikaevskii
equation results is quite good.
For these experimental runs, the length of the first pulse was adjusted 
such that half the population was in the state with zero momentum 
and half in the state with finite momentum. We emphasize that the
resulting  pulse length of
$0.207$~ms does not correspond to a 50~\% population transfer in the absence
of interactions and vanishing momentum spread along the $z$-direction
of the initial state. The second pulse was taken to be $0.200$~ms.
To calibrate the coupling strength, we performed calculations for 
different $\Omega_{0,L}$ and picked the value that yields, using a $0.207$~ms
pulse, a 50/50 population distribution after the first pulse.

While the agreement between the symbols and solid lines
in Fig.~\ref{fig_ramseylatticeexp} is
not perfect, the theoretical and experimental data
share several key characteristics:
(i)
The number of fringes increases with increasing hold time.
(ii)
The density pattern is not characterized by a single fringe spacing; rather, the fringe
spacings seem to vary across the expanded cloud. 
(iii)
The density displays a small amplitude for
$z$-values around $-75$~$\mu$m and 
$150$~$\mu$m; these peaks correspond to momentum 
space components centered
around $-2 \hbar k_L$ and $4 \hbar k_L$, respectively. 
(iv)
The density distributions centered around $z \approx 0$ (corresponding
to the component with momentum along the $z$-direction of $\approx 0$)
and centered around $z \approx 75$~$\mu$m
(corresponding
to the component with momentum of $\approx 2 \hbar k_L$)
have fairly distinct shapes, i.e., they are not mirror images 
of each other.
All these observations are
consistent with the discussion presented in the previous
section. If the interaction effects played less of a role, the interference
pattern would be ``cleaner", i.e., more regular.

As already alluded to earlier, Ref.~\cite{phillips2000} measured the linear and quadratic
phases using a Ramsey-type Bragg pulse sequence. Their analysis assumed
equally spaced fringes.
While the fringe pattern in Fig.~2 of Ref.~\cite{phillips2000}
is more ``regular" than the fringe pattern 
displayed in Fig.~\ref{fig_ramseylatticeexp}, the density peaks in 
Fig.~2(f) of Ref.~\cite{phillips2000} are, just as in our case, not fully
symmetric with respect to the midpoint. We speculate that this might 
be
due to the structural dynamics that is driven by mean-field effects
(``factor of 2")
discussed in our work for the lattice coupling case.

While the overall agreement between the experimental and theoretical
data in Fig.~\ref{fig_ramseylatticeexp}
is satisfactory, the experimental data hint at the presence of beyond
mean-field physics. In particular, we consistently observe a significant fraction of atoms 
``between" the two clouds, i.e., with a momentum of around $\hbar k_L$.
It is presently unclear if this is due to quantum correlations that are not captured by the
mean-field Gross-Pitaevskii equation or if, possibly, the thermal cloud plays a non-negligible role.
A detailed investigation of these questions is beyond the scope of this work.

\section{Summary and outlook}
\label{sec_outlook}

This paper 
investigated
two realizations of
a two-state model; in both realizations, the two states 
are represented by 
a spatially- and time-dependent mean-field wave function or orbital.
The description goes beyond a class of simpler mean-field models, where the
dynamics of each mode is described by one complex number that encodes
the population and phase of the mode, thereby assuming that 
the spatial
dynamics of the modes play a negligible 
role~\cite{work_on_two_mode_c_number1,work_on_two_mode_c_number2,work_on_two_mode_c_number3}.
Our
work demonstrates that time-dependent deformations of the spatial profile of the
mean-field wave functions play an important role
when the two-state model is realized by loading
a single-component $^{87}$Rb BEC into a moving one-dimensional optical lattice
that introduces a coupling between two distinct
momentum states of the atom. 
When the two-state model
is, instead, realized by coupling two different
hyperfine states of $^{87}$Rb BEC atoms through a two-photon Raman process,
time-dependent deformations of the spatial profile of the
mean-field wave functions are notably less pronounced.

The difference in the dynamics for the two physical realizations (lattice and Raman coupling, respectively)
of the two-mode model
was
traced back to the contribution of different scattering diagrams;
in particular,
there exist two scattering diagrams 
(these are depicted in the second row in Fig.~\ref{fig_collision})
that contribute in the lattice coupling case
but not in the Raman coupling case (due to the ``factor of 2"). 
Said differently, the 
mean-field interactions for the lattice and Raman coupling cases differ:
The effective two-state model for the lattice
coupling case contains two repulsive ``off-site" interaction
terms 
that are absent in the Raman coupling case.  
As a consequence, the lattice coupled system
is characterized by an
enhanced tendency for phase
separation, which ``competes" with the lattice coupling term
that has a tendency to keep the components together.
This competition gives rise to
the internal mean-field dynamics in the lattice coupled system
that is different from
the mean-field dynamics displayed by
the two-state Hamiltonian for the Raman coupling
case.

While the discussion throughout this paper focused on $^{87}$Rb
BECs, the lattice coupling results, which rely
on the occupation of a single hyperfine state, apply to any  
BEC with positive two-body $s$-wave scattering length.
The Raman coupling results were obtained assuming that
the four coupling strengths 
$g_{aa}$, $g_{bb}$, $g_{ab}$, and  $g_{ba}$ are approximately equal or equal to each other; 
this assumption holds for the $F=1$ states
of $^{87}$Rb but not necessarily
for other elements.
  
 The present work has a number of
 practical and conceptual
 implications:
\begin{itemize}
\item The Rabi oscillation data (see Figs.~\ref{fig_ramanrabi}, \ref{fig_ramanrabiexp},
\ref{fig_lattice_rabi}, and \ref{fig_latticerabiexp}) show,
especially for weak coupling strengths, pronounced
non-sinusoidal behavior. This indicates that the analysis of experimental
Rabi oscillation data, taken to calibrate the coupling strength, 
has to proceed with care. A simple fit to a sinusoidal function 
(or damped sinusoidal function) may yield an imprecise
coupling strength due to interaction effects. Such data can be 
used for calibration purposes if compared with mean-field simulations that account for the interaction effects. Alternatively, experiments can operate in the dilute regime
where interaction effects are negligible.
Related discussions of lattice potential calibrations can be found in 
Refs.~\cite{latticeGP,cristiani2002,lattice_calibration,gardiner2019}.
\item In the ``weak" lattice coupling case---this is the
regime where, as discussed in Sec.~\ref{sec_theory_lattice}, the effective
two-state model Hamiltonian provides a reliable description
of the system dynamics---the internal dynamics leads to a deformation of
the density profiles of the components.
We 
argued
that these density deformations can be interpreted as corresponding
to an effective
position-dependent detuning.
For example, starting with all population
in one of the two states, a $\pi/2$-pulse (defined for a
single atom in free space), realized using a comparatively weak coupling strength, 
yields a state with a population distribution
that differs from a 50/50 mixture. This fact, together with the
build-up of spatial deformations during the hold time, has implications for 
momentum space engineering protocols, which aim to
implement beam-splitters and other operations that are commonly
realized in quantum optics~\cite{edwards2010,edwards2011}.
\item
Integrating out the spatial degrees of freedom,
the dynamics of the two-state Hamiltonian considered in this
work 
reduces to coupled mean-field equations
that are characterized by two complex numbers,
representing the populations and phases 
of the two modes~\cite{work_on_two_mode_c_number1,work_on_two_mode_c_number2,work_on_two_mode_c_number3,arimondo2003}.
In the lattice case, this reduced dimensionality model
has been shown to support intriguing 
swallow-tail lattice 
structures~\cite{work_on_two_mode_c_number1,swallowtail1,swallowtail3,swallowtail4,swallowtail5}, 
which support, e.g., 
mean-field induced non-exponential tunneling~\cite{work_on_two_mode_c_number1,work_on_two_mode_c_number3}.
The internal spatial dynamics highlighted in the present work suggests that 
the validity regime of these reduced dimensionality models needs to be
assessed carefully.
\end{itemize}

\section{Acknowledgement}
\label{acknowledgement}
We thank Vandna Gokhroo for her contributions during the initial stage of this project.
Support by the National Science Foundation through
grant numbers
PHY-1806259 (QG and DB), PHY-1607495 (TMB, SM, and PE), and PHY-1912540 (TMB, SM, and PE) are
gratefully acknowledged.
This work used
the OU
Supercomputing Center for Education and Research
(OSCER) at the University of Oklahoma (OU).


\begin{thebibliography}{100}

\bibitem{quantumtext}
D. J. Griffiths, Introduction to Quantum Mechanics (Pearson Prentice Hall, 2005).

\bibitem{eberly}
L. Allen and J. H. Eberly, Two-level Atoms and Optical Resonance (Dover, 1987).

\bibitem{rabi1939}
I. I. Rabi, S. Millman, P. Kusch, and J. R. Zacharias, 
The molecular beam resonance method for measuring nuclear magnetic moments,
Phys. Rev. {\bf{55}}, 526 (1939).

\bibitem{saffman2008}
T. A. Johnson, E. Urban, T. Henage, L. Isenhower, D. D. Yavuz, T. G. Walker, and M. Saffman,
Rabi Oscillations between Ground and Rydberg States with Dipole-Dipole Atomic Interactions,
Phys. Rev. Lett. {\bf{100}}, 113003 (2008). 

\bibitem{gambarelli2012}
H. De Raedt, B. Barbara, S. Miyashita, K. Michielsen, S. Bertaina, and S. Gambarelli,
Quantum simulations and experiments on Rabi oscillations of spin qubits: 
Intrinsic vs extrinsic damping,
Phys. Rev. B {\bf{85}}, 014408 (2012).

\bibitem{chen2019}
C.-H. Li, C. Qu, R. J. Niffenegger, S.-J. Wang, M. He, D. B. Blasing, A. J. Olson, 
C. H. Greene, Y. Lyanda-Geller, Q. Zhou, C. Zhang, Y. P. Chen,
Spin current generation and relaxation in a quenched spin-orbit-coupled Bose-Einstein condensate,
Nat. Comm. {\bf{10}}, 375 (2019).

\bibitem{spielman2011}
Y.-J. Lin, K. Jim{\'{e}}nez-Garc{\'{i}}a and I. B. Spielman, 
Spin-orbit-coupled Bose-Einstein condensates,
Nature {\bf{471}}, 83 (2011).

\bibitem{spielman2013}
V. Galitski and I. B. Spielman, 
Spin-orbit coupling in quantum gases,
Nature {\bf{494}}, 49 (2013).

\bibitem{zhai2015}
H. Zhai,
Degenerate quantum gases with spin-orbit coupling: a review,
Rep. Prog. Phys. {\bf{78}}, 026001 (2015).


\bibitem{salomon1996}
M. B. Dahan, E. Peik, J. Reichel, Y. Castin, and C. Salomon,
Bloch Oscillations of Atoms in an Optical Potential,
Phys. Rev. Lett. {\bf{76}}, 4508 (1996).


\bibitem{phillips1999}
M. Kozuma, L. Deng, E. W. Hagley, J. Wen, R. Lutwak, K. Helmerson, S. L. Rolston, and W. D. Phillips,
Coherent Splitting of Bose-Einstein Condensed Atoms with Optically Induced Bragg Diffraction,
Phys. Rev. Lett. {\bf{82}}, 871 (1999).

\bibitem{morsch2001}
O. Morsch, J. H. M{\"u}ller, M. Cristiani, D. Ciampini, and E. Arimondo,
Bloch Oscillations and Mean-Field Effects of Bose-Einstein Condensates in 1D Optical Lattices,
Phys. Rev. Lett. {\bf{87}}, 140402 (2001).


\bibitem{ramsey1985}
N. Ramsey, 
Molecular Beams (Clarendon, 1985).

\bibitem{castin1996}
Y. Castin and R. Dum,
Bose-Einstein Condensates in Time Dependent Traps,
Phys. Rev. Lett. {\bf{77}}, 5315 (1996). 


\bibitem{other_castin}
P. B. Blakie and R. J. Ballagh, 
Mean-field treatment of Bragg scattering from a Bose-Einstein condensate, 
J. Phys. B {\bf{33}}, 3961 (2000).

\bibitem{gupta2011}
A. O. Jamison, J. N. Kutz, and S. Gupta,
Atomic interactions in precision interferometry using Bose-Einstein condensates,
Phys. Rev. A {\bf{84}}, 043643 (2011).



\bibitem{edwards2010}
M. Edwards, B. Benton, J. Heward, and C. W. Clark,
Momentum-space engineering of gaseous Bose-Einstein condensates,
Phys. Rev. A {\bf{82}}, 063613 (2010).


\bibitem{gadway2015}
B. Gadway,
Atom-optics approach to studying transport phenomena,
Phys. Rev. A {\bf{92}}, 043606 (2015).

\bibitem{experiment_soc}
C. Qu, C. Hamner, M. Gong, C. Zhang, and P. Engels,
Observation of {\em{Zitterbewegung}} in a spin-orbit-coupled Bose-Einstein condensate,
Phys. Rev. A {\bf{88}}, 021604(R) (2013).



\bibitem{experiment_optical_lattice}
C. Hamner, Y. Zhang, M. A. Khamehchi, M. J. Davis, and P. Engels,
Spin-Orbit-Coupled Bose-Einstein Condensates in a One-Dimensional Optical Lattice,
Phys. Rev. Lett. {\bf{114}}, 070401 (2015).


\bibitem{scatteringlength_rb}
The values of the scattering lengths are taken from
M. A. Khamehchi, Y. Zhang, C. Hamner, T. Busch, and P. Engels,
Measurement of collective excitations in a spin-orbit-coupled Bose-Einstein condensate,
Phys. Rev. A {\bf{90}}, 063624 (2014).

\bibitem{chebychev1}
  H.~Tal-Ezer, and R.~Kosloff, 
  An accurate
  and efficient scheme for propagating the time dependent Schr{\"o}dinger
  equation,
  J. Chem. Phys. {\bf{81}}, 3967 (1984).
 
\bibitem{chebychev2}
 C. Leforestier, R.~H. Bisseling, C.~Cerjan,
 M.~D. Feit,  R.~Friesner, 
  A.~Guldberg, 
 A.~Hammerich, 
G.~Jolicard, 
 W.~Karrlein, 
  H.~D. Meyer, 
  N. Lipkin, O. Roncero, and R. Kosloff,
 A Comparison of Different Propagation
  Schemes for the Time Dependent Schr{\"o}dinger Equation,
  J. Comput. Phys. {\bf{94}}, 59 (1991).
  
                           

\bibitem{stringariRMP}
F. Dalfovo, S. Giorgini, L. P. Pitaevskii, and S. Stringari,
Theory of Bose-Einstein condensation in trapped gases,
Rev. Mod. Phys. {\bf{71}}, 463 (1999).


\bibitem{phillips2000}
J. E. Simsarian, J. Denschlag, M. Edwards, C. W. Clark, 
L. Deng, E. W. Hagley, K. Helmerson, S. L. Rolston, and W. D. Phillips,
Imaging the Phase of an Evolving Bose-Einstein Condensate Wave Function,
Phys. Rev. Lett. {\bf{85}}, 2040 (2000).



\bibitem{edwards2011}
B. Benton, M. Krygier, J. Heward, M. Edwards, and C. W. Clark,
Prototyping method for Bragg-type atom interferometers,
Phys. Rev. A {\bf{84}}, 043648 (2011).


\bibitem{bongs2001}
K. Bongs, S. Burger, S. Dettmer, D. Hellweg, J. Arlt, W. Ertmer, and K. Sengstock,
Waveguide for Bose-Einstein condensates,
Phys. Rev. A {\bf{63}}, 031602(R) (2001).

\bibitem{latticeGP}
O. Morsch and M. Oberthaler,
Dynamics of Bose-Einstein condensates in optical lattices,
Rev. Mod. Phys. {\bf{78}}, 179 (2006).



\bibitem{phase_separation}
E. Timmermans,
Phase Separation of Bose-Einstein Condensates,
Phys. Rev. Lett. {\bf{81}}, 5718 (1998).


\bibitem{chin2013}
C. V. Parker, L.-C. Ha, and C. Chin, 
Direct observation of effective ferromagnetic domains of cold atoms in a shaken optical lattice,
Nat. Phys. {\bf{9}}, 769 (2013).

\bibitem{gadway_lattice}
F.  A. An, E. J. Meier, J. Ang'ong'a, and B. Gadway,
Correlated Dynamics in a Synthetic Lattice of Momentum States,
Phys. Rev. Lett. {\bf{120}}, 040407 (2018).

\bibitem{work_on_two_mode_c_number1}
B. Wu and Q. Niu,
Nonlinear Landau-Zener tunneling,
Phys. Rev. A {\bf{61}}, 023402 (2000).

\bibitem{work_on_two_mode_c_number2}
O. Zobay and B. M. Garraway,
Time-dependent tunneling of Bose-Einstein condensates,
Phys. Rev. A {\bf{61}}, 033603 (2000).

\bibitem{work_on_two_mode_c_number3}
J. Liu, L. Fu, B.-Y. Ou, S.-G. Chen, D.-I. Choi, B. Wu, and Q. Niu,
Theory of nonlinear Landau-Zener tunneling,
Phys. Rev. A {\bf{66}}, 023404 (2002).


\bibitem{cristiani2002}
M. Cristiani, O. Morsch, J. H. M\"uller, D. Ciampini, and E. Arimondo,
Experimental properties of Bose-Einstein condensates in
one-dimensional optical lattices:
Bloch oscillations, Landau-Zener tunneling, and mean-field effects,
Phys. Rev. A {\bf{65}}, 063612 (2002).


\bibitem{lattice_calibration}
C. Cabrera-Guti{\'e}rrez, E. Michon, V. Brunaud, T. Kawalec, A. Fortun, M. Arnal, J. Billy, and 
D. Gu{\'e}ry-Odelin,
Robust calibration of an optical-lattice depth based on a phase shift,
Phys. Rev. A {\bf{97}}, 043617 (2018).

\bibitem{gardiner2019}
B. T. Beswick, I. G. Hughes, and S. A. Gardiner,
Lattice-depth measurement using continuous grating atom diffraction,
Phys. Rev. A {\bf{100}}, 063629 (2019).

\bibitem{arimondo2003}
M. Jona-Lasinio, O. Morsch, M. Cristiani, N. Malossi, J. H. M\"uller, E. Courtade,
M. Anderlini, and E. Arimondo,
Asymmetric Landau-Zener Tunneling in a Periodic Potential,
Phys. Rev. Lett. {\bf{91}}, 230406 (2003).



\bibitem{swallowtail1}
K. Berg-S{{\o}}rensen and K. M{\o}lmer,
Bose-Einstein condensates in spatially periodic potentials,
Phys. Rev. A {\bf{58}}, 1480 (1998).


\bibitem{swallowtail3}
B. Wu, R. B. Diener, and Q. Niu,
Bloch waves and bloch bands of Bose-Einstein condensates in optical lattices,
Phys. Rev. A {\bf{65}}, 025601 (2002).

\bibitem{swallowtail4}
D. Diakonov, L. M. Jensen, C. J. Pethick, and H. Smith,
Loop structure of the lowest Bloch band for a Bose-Einstein condensate,
Phys. Rev. A {\bf{66}}, 013604 (2002).

\bibitem{swallowtail5}
M. Machholm, C. J. Pethick, and H. Smith,
Band structure, elementary excitations, and stability of a Bose-Einstein condensate in a periodic potential,
Phys. Rev. A {\bf{67}}, 053613 (2003).






\end{thebibliography}
\end{document}